\documentclass[12pt]{article}
\usepackage[utf8]{inputenc}
\usepackage[a4paper,margin=3cm]{geometry}
\usepackage{amsmath}
\usepackage{amssymb}
\usepackage{bm}
\usepackage{subcaption}
\usepackage{graphicx}
\usepackage{cite}
\usepackage{url}
\usepackage{xcolor, soul}
\usepackage{hyperref}
\colorlet{clear}{yellow!0}
\sethlcolor{clear} 

\begin{document}

\title{Scaling laws for electron kinetic effects in tokamak scrape-off layer plasmas}
\author{D Power$^1$\footnote{d.power19@imperial.ac.uk} \and S Mijin$^2$ \and M Wigram$^3$ \and F Militello$^2$ \and R Kingham$^1$}
\date{%
    $^1$Blackett Lab., Plasma Physics Group, Imperial College London, London, United Kingdom\\%
    $^2$UKAEA, Culham Science Centre, Oxon, United Kingdom\\%
    $^3$MIT Plasma Science and Fusion Center, Cambridge, MA 02139, USA\\[2ex]%
    \today
}

\maketitle

\begin{abstract}
    Tokamak edge (scrape-off layer) plasmas can exhibit non-local transport in the direction parallel to the magnetic field due to steep temperature gradients. This effect along with its consequences has been explored at equilibrium for a range of conditions, from sheath-limited to detached, using the 1D kinetic electron code SOL-KiT, where the electrons are treated kinetically and compared to a self-consistent fluid model. Line-averaged suppression of the kinetic heat flux (compared to Spitzer-H\"{a}rm) of up to {50\%} is observed, contrasting with up to {98\%} enhancement of the sheath heat transmission coefficient, $\gamma_e$. Simple scaling laws in terms of basic SOL parameters for both effects are presented. By implementing these scalings as corrections to the fluid model, we find good agreement with the kinetic model for target electron temperatures. 
    It is found that the strongest kinetic effects in $\gamma_e$ are observed at low-intermediate collisionalities, and tend to increase at increasing upstream densities and temperatures. On the other hand, the heat flux suppression is found to increase monotonically as upstream collisionality decreases. {The conditions simulated encompass collisionalities relevant to current and future tokamaks.}
\end{abstract}

\clearpage

\section{Introduction}

The region of unconfined plasma at the edge of tokamaks, called the scrape-off layer (SOL), is the barrier between the hot core plasma and the solid surfaces which make up the inside of the reactor. It is necessary to understand plasma transport in this region, which occurs primarily parallel to the magnetic field lines, so that accurate predictions can be made for future devices and steps can be taken to mitigate heat fluxes which may exceed material constraints.

Transport in SOL plasmas is often treated with fluid models, where a Braginskii-like set of transport equations \cite{Braginskii2004} may be solved. However, the presence of steep temperature gradients parallel to the magnetic field, as would be expected in reactor-class devices, means heat transport (particularly for the electrons) {is dominated by particles with collisional mean free paths which are long relative to the temperature gradient length scale} and so becomes `non-local'.
This can be quantified with the upstream collisionality parameter $\nu_{u}^* = L/\lambda_{u}$ \cite{Stangeby2001},
defined as the ratio of the parallel SOL length $L$ and the upstream mean free path $\lambda_u$. Conditions where $\nu_{u}^*$ is small and temperature gradients are large may not be described accurately by a fluid model.

This effect has been explored in recent years \cite{Mijin2020a,Wigram2020,Zhao2019,Chankin2018,Brodrick2017,Tskhakaya2015,Omotani2013,Batishchev1999,AbouAssaleh1994}, where it is now well-documented that kinetic suppression of the heat flux can result in steeper temperature gradients and lower target temperatures when compared to a fluid model. Somewhat less understood is the region of operating parameter space where such effects may become important, and the consequences for the overall energy balance at equilibrium (i.e. how energy going into the SOL makes its way out). It is still unclear whether kinetic effects in parallel transport pose a significant uncertainty in modelling approaches for future devices. 

Here we present kinetic and fluid simulations of a 1D SOL plasma model, across a wide range of the relevant parameter space (input power, plasma density {and connection length}), in order to assess and understand kinetic deviations from fluid model predictions. The model will briefly be presented in Section \ref{sec:SOL-KiT}, followed by an explanation of the simulations that have been carried out in Section \ref{sec:simulations}. We will then summarise the results (Section \ref{sec:results}), highlighting the areas in which kinetic effects are (and are not) seen. Following a discussion of the observed results (Section \ref{sec:discussion}), we present scaling relationships in terms of basic SOL parameters for the main kinetic effects seen - enhancement to the sheath heat transmission coefficient and suppression of the parallel conductive electron heat flux - in Section \ref{sec:predictions}. These will be used to reproduce electron temperature profiles from kinetic simulations in a (corrected) fluid model. {A prediction is made for the strength of these kinetic effects in ITER.} 

\section{Kinetic and fluid modelling with SOL-KiT}
\label{sec:SOL-KiT}

SOL-KiT is a fully implicit 1D plasma code which has been used to study kinetic effects in parallel electron transport in the SOL \cite{Mijin2019,Mijin2020a,Power2021}. Here a very brief outline of SOL-KiT is presented, and the reader is referred to \cite{Mijin2021} for more details of the model. 

In kinetic mode, SOL-KiT solves the VFP equation for electrons in a hydrogenic plasma along the direction parallel to the magnetic field (the $x$-axis), 

\begin{equation}\label{eq:vfp}
    \frac{\partial f(x, \vec{v}, t)}{\partial t}+v_{x}\frac{\partial f(x, \vec{v}, t)}{\partial x} -\frac{e}{m_{e}} E \frac{\partial f(x, \vec{v}, t)}{\partial v_{x}}=\sum_{\alpha} C_{e-\alpha},
\end{equation}

where $f(x, \vec{v}, t)$ is the electron velocity distribution, which is a function of space, velocity and time. $E$ is the electric field along $x$, $m_e$ is the electron mass, $-e$ is the electron charge and $v_x$ is the electron velocity along $x$. The right hand side consists of Fokker-Planck collisions (electron-electron and electron-ion) and Boltzmann collisions (electron-neutral). A spherical harmonic decomposition in velocity space is used to solve this equation as outlined in \cite{shkarofsky1966}. Azimuthal symmetry is assumed about the $x$-axis so that the magnetic field may be ignored.

The $x$-axis spans from the midplane at $x=0$ (`upstream'), to the plasma sheath boundary at $x=L$, where $L$ is the domain length, which is half the connection length (the parallel distance between two strike points in a divertor SOL). Power enters the plasma upstream, and leaves at the sheath or through plasma-neutral collisions. The included collisional processes are electron-impact ionisation and excitation; the inverse of these processes (three-body recombination and collisional de-excitation); and resonant charge exchange (CX) between the ions and neutrals. In addition, radiative recombination and de-excitation are modelled, which allows energy to leave the plasma-neutral system.  

In fluid mode, moments of equation (\ref{eq:vfp}) are solved instead, allowing for a direct comparison between a fluid and kinetic treatment. Evolution equations for the electron temperature $T_e$, flow velocity $u_e$ and density $n_e$ are solved. The $T_e$ equation is closed with Braginskii/Spitzer-H\"{a}rm heat flow \cite{Braginskii2004,Spitzer1953}, $q_{\parallel,e}=-\kappa_{e} \frac{\partial\left(k T_{e}\right)}{\partial x}+0.71 n_{e} k T_{e}\left(u_{e}-u_{i}\right)$.

Quasi-neutrality is enforced by setting the ion density $n_i=n_e$. The parallel electric field $E$ is evolved with Amp\`{e}re-Maxwell's law, $\frac{\partial E}{\partial t} = -(j_e - j_i)/\epsilon_0$, where the ion and electron currents are $j_{i,e}=\pm e n_{i,e}u_{i,e}$. The implicit time integration used by SOL-KiT means this results in ambipolarity when using timesteps large relative to the plasma oscillation period.

In kinetic mode, where fluid electron quantities are required, for example $u_e$ in calculating $E$, the appropriate velocity moments of $f$ are taken.

In order to provide a realistic background on which to solve electron transport, SOL-KiT also models the hydrogenic ions and neutral atoms, both of which are treated with fluid models (discussed more in the next section). Atomic processes including ionisation, recombination, excitation and de-excitation are handled by solving a collisional-radiative model (CRM) for the neutral atoms alongside the transport equations, using fundamental cross-sections and rates from Janev \cite{Janev2003} and NIST \cite{Kramida2020}. The fully time-dependent CRM coupled with neutral transport gives a non-coronal model. Particle sources and the effect of electron-neutral collisions on the electrons are evaluated with the inelastic Boltzmann collision operator. In fluid mode, the kinetic collision operators are evaluated with a Maxwellian electron distribution. At present no impurity species are treated by SOL-KiT, although this is planned for a future update to the code. 

The upstream boundary is reflective. At the sheath boundary, the Bohm criterion is applied so that plasma flow reaches the sound speed $c_s$. All plasma particle flux across this boundary is lost and recycled as atomic neutrals, which are placed in the last spatial cell. In kinetic mode, the logical boundary condition \cite{Procassini1992} is applied to the electron distribution, where the forward-going part of the distribution is reflected and truncated at some velocity $v_c$, which is calculated iteratively to ensure equal electron and ion fluxes. In fluid mode, this same boundary condition manifests via the heat flux at the sheath entrance, $q_{sh,e}=\gamma_e kT_{e,t} \Gamma_t$, for target temperature $T_{e,t}$ and particle flux $\Gamma_t$, and where 
\begin{equation}
    \label{eq:gamma_e_def}
\gamma_e=2-0.5 \ln \left(2 \pi\left(1+T_{i,t} / T_{e,t}\right) m_{e} / m_{i}\right)
\end{equation}
is the sheath heat transmission coefficient \cite{Stangeby2001}.

\subsection{Extensions to SOL-KiT}

For this study, this SOL-KiT model described in \cite{Mijin2021} has been extended to provide a more realistic background plasma on which to solve the electron transport, as well as to provide some reduction in compute time. As some of these improvements have not yet been documented elsewhere, they are outlined here.

\subsubsection{Ion temperature equation}


Firstly, an ion temperature equation has been added to the model, allowing for ion energy transport independently from the electrons as well as additional channels for ion-electron and ion-neutral energy transfer. More detail on this change is provided in \cite{Power2021}, omitted here for brevity. The ion heat flow is $q_{\parallel,i}=-\kappa_{i} \frac{\partial\left(k T_{i}\right)}{\partial x}$, using the Spitzer-H\"{a}rm $\kappa_{i}$ \cite{Spitzer1953}. The sheath boundary condition on the ion temperature equation is equivalent to the one for the fluid electron model, $q_{sh,i}=\gamma_i kT_{i,t} \Gamma_t$, with $\gamma_i=2.5$ {\mbox{\cite{Stangeby2001}}}. In fluid mode, ion-electron energy transfer is treated with $Q_{ie}=-Q_{ei}=-\frac{3 m_{e}}{m_{i}} \frac{n_{e} k}{\tau_{e}}\left(T_{i}-T_{e}\right)$ \cite{Braginskii2004}, where $\tau_e$ is the electron collision time. In kinetic mode, we instead take the energy moment of the Fokker-Planck collision operator assuming Maxwellian ions, $Q_{ei}=\int \frac{1}{2}m_ev^2C_{ei}d\vec{v}$. The two approaches are equivalent for Maxwellian electrons \cite{Power2021}.

\begin{figure}
  \centering 
  \includegraphics[width=0.5\textwidth]{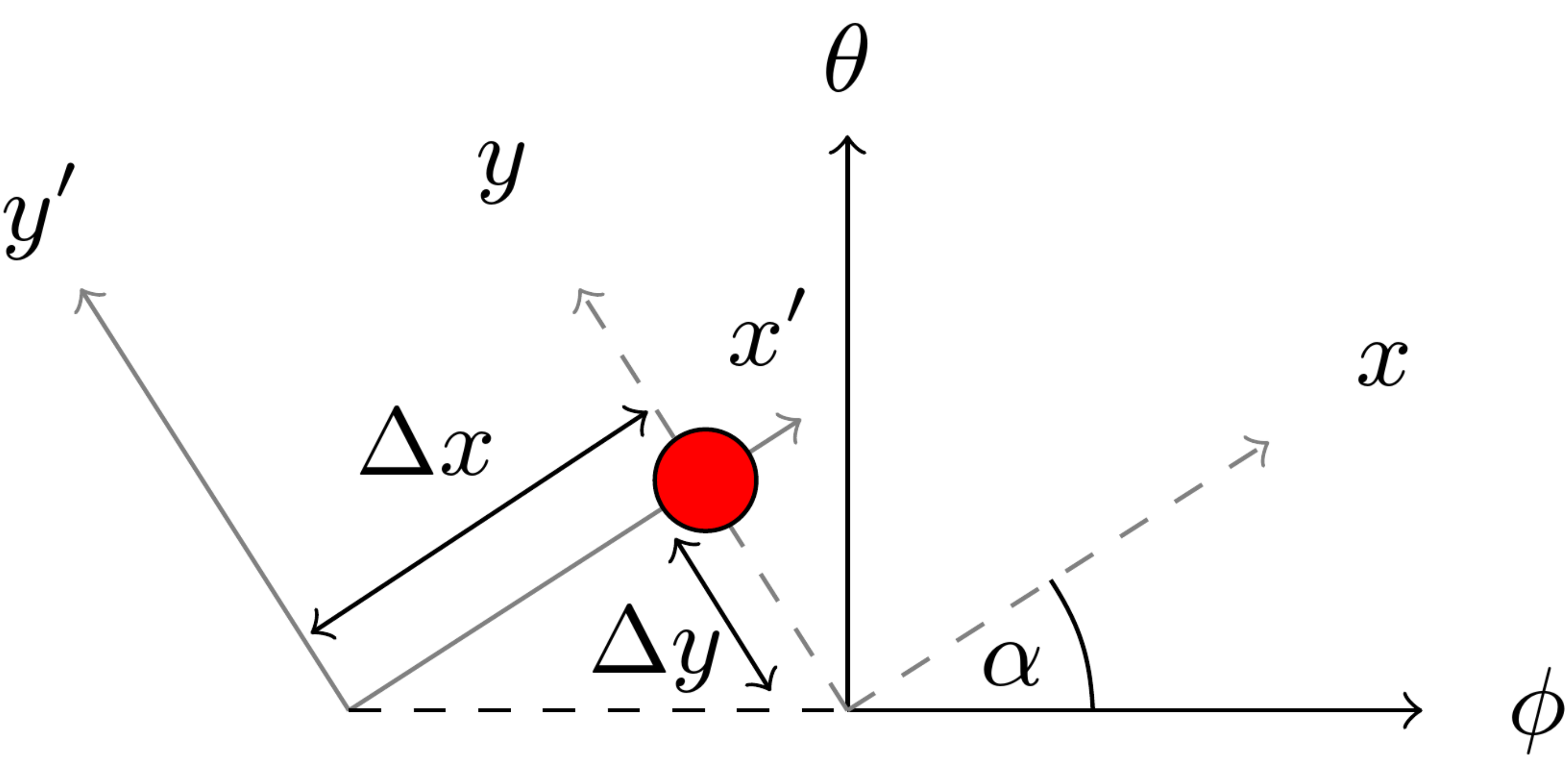}
  \caption{{Geometry of the quasi-2D neutral transport model; $x$ and $y$ are the parallel and perpendicular directions respectively; $\phi$ and $\theta$ are the toroidal and poloidal directions; $\alpha$ is the pitch angle of the magnetic field. In an $x-y$ coordinate system, assuming toroidal symmetry, $y$ coordinates are transformed into $x$ coordinates by translation in the $\phi$ direction (shown by the $x^\prime-y^\prime$ axes).}}
  \label{fig:quasi2d_geometry}
\end{figure}

\subsubsection{Fluid neutral model}

Secondly, velocity and temperature equations have been added for the (hydrogenic) neutral particles, in order to capture the important processes of energy and momentum transfer between ions and neutrals via particle exchange from ionisation, recombination and CX. 
{
  In order to capture neutral transport in the direction perpendicular to the magnetic field lines, we use the assumption of toroidal symmetry to project perpendicular velocities back on to the parallel direction\mbox{\cite{Blommaert2018,Dudson2019}}. The geometry is shown in Figure \mbox{\ref{fig:quasi2d_geometry}}. For a velocity $\vec{u}= u_\parallel \hat{x} + u_\perp \hat{y}$, where $\hat{y}$ is the unit vector perpendicular to the magnetic field, we can obtain an effective velocity in the parallel direction,
  $u_{eff}=u_{\parallel} + u_{\perp}/\tan \alpha$, where $\alpha = \tan^{-1}(B_\theta / B_\phi)$ is the pitch angle of the magnetic field lines relative to the wall. Similarly, we can relate perpendicular gradients to parallel gradients via $\frac{\partial}{\partial y} = \frac{1}{\tan \alpha} \frac{\partial}{\partial x}$. With this, we obtain a 1D neutral transport model which captures something of the cross-field neutral transport. The parallel neutral velocity equation is
}
\begin{equation}\label{eq:u_n_par}
    \begin{split}
        \frac{\partial u_{n \|}}{\partial t}=-u_{n, e f f} \frac{\partial u_{n \|}}{\partial x}-\frac{1}{m_{n} n_{n}} \frac{\partial p_{n}}{\partial x} -\frac{S_{n}}{n_{n}} u_{n \|}+\frac{1}{m_{n} n_{n}} R_{n \|},
    \end{split}
\end{equation}
{and the perpendicular velocity equation is}
\begin{equation}\label{eq:u_n_perp}
    \begin{split}
        \frac{\partial u_{n \perp}}{\partial t}=-u_{n, e f f} \frac{\partial u_{n \perp}}{\partial x}-\frac{1}{m_{n} n_{n} \tan \alpha} \frac{\partial p_{n}}{\partial x}-\frac{S_{n}}{n_{n}} u_{n \perp}+\frac{1}{m_{n} n_{n}} R_{n \perp},
    \end{split}
\end{equation}
where $p_n=n_nkT_n$ is the neutral pressure, $S_n$ is the neutral particle source and $R_{n,\parallel}$, $R_{n,\perp}$ are the parallel and perpendicular momentum sources. 
{These contain contributions from particle sources and CX. We use a similar form of the CX friction term outlined by Meier\mbox{\cite{Meier2011}}, along with velocity-dependent CX cross-sections from Janev\mbox{\cite{Janev2003}}.}
For calculating $R_{n,\perp}$, the plasma is assumed to be stationary in the perpendicular direction. 
{The pitch angle $\alpha$} in practice determines the degree to which neutrals are able to transport upstream in this 1D model. {For normal incidence, $\alpha=90^\circ$, $u_{n,eff}=u_{n\parallel}$ and the model is 1D along $x$.}
{The connection of this model with the pressure-diffusion neutral model in\mbox{\cite{Horsten2017}}, and a similar quasi-2D transport model implemented in SD1D\mbox{\cite{Dudson2019}}, can be seen by dropping all but the pressure gradient and friction terms in (\mbox{\ref{eq:u_n_perp}}). The perpendicular velocity is then $u_{n,\perp}\propto \frac{1}{\tan \alpha} \frac{\partial p_n}{\partial x}$, which can be inserted into $u_{n,eff}$ to yield a convective-diffusive model, where the diffusion coefficient is enhanced by a geometric factor proportional to $1/\tan^2 \alpha =(B_\phi/B_\theta)^2$.}

SOL-KiT solves a continuity equation for each excited state of a neutral hydrogenic species, denoted with subscript $b$, each with its own density $n_b$ and particle source $S_b$. Total neutral density is therefore $n_n = \sum_b n_b$, and $S_n=\sum_b S_b$. The additional velocity equations (\ref{eq:u_n_par} \& \ref{eq:u_n_perp}) yield a modified continuity equation for each species,
\begin{equation}
    \frac{\partial n_{b}}{\partial t}=-\frac{\partial\left(n_{b} u_{n, e f f}\right)}{\partial x}+S_{b}.
\end{equation}

The neutral temperature $T_n$ is also evolved,
\begin{equation}\label{eq:T_n}
    \begin{aligned}
      \frac{\partial k T_{n}}{\partial t}=-u_{n,eff} \frac{\partial T_{n}}{\partial x}+\frac{2}{3}\left[\frac{Q_{n}}{n_{n}}\right. - & \left. k T_{n} \frac{\partial u_{n,eff}}{\partial x}-\frac{S_{n}}{n_{n}}\left(\frac{3}{2} k T_{n}-\frac{1}{2} m_{n} u_{n \|}^{2}\right) \right. \\
      & - \left.\frac{1}{n_{n}}\left(1+\frac{1}{\tan ^{2} \alpha}\right) \frac{\partial q_{n}}{\partial x}-\frac{u_{n \|} R_{n \|}}{n_{n}}\right],
  \end{aligned}
\end{equation}
where $q_n$ is the neutral heat flow, $q_{n}=-2.4\left(\frac{n_{n} T_{n}}{m_{n} \nu_{C X}}\right)\frac{\partial T_{n}}{\partial x}$ \cite{Helander1994} and $\nu_{CX}$ is the CX collision frequency. The energy source term $Q_n$ contains contributions from CX collisions and the energy transfer from particle exchange during ionisation and recombination. {As in the case of CX friction, for the CX contribution to $Q_n$ we use a similar form to that in\mbox{\cite{Meier2011}}.} 

{The boundary condition on the fluid neutral model is a simplified version of the approach outlined by Horsten et al.\mbox{\cite{Horsten2016,Horsten2017}}, where the neutral distribution at the boundary is taken to be made up of a set of one-sided distributions from the incident, reflected and recycled neutrals. We assume 100\% reflection of the neutrals incident on the wall and no `fast' recycling of ions, i.e. they are thermally released from the wall as molecules which quickly dissociate. The ion particle flux crossing the sheath, $\Gamma_{i,t}$, is recycled and re-enters as a neutral flux at the same location. These recycled particles are assumed to be emitted from the wall mono-energeticaly at the Frank-Condon dissocation energy $T_{FC}=3$eV. Neutrals incident on the wall are reflected back with some loss of energy. The net energy flux of neutrals at the boundary is therefore (for 100\% plasma recycling) $q_{n,t}=\gamma_nkT_n n_n c_{n,s} - kT_{FC} \Gamma_{i,t}$, where $c_{n,s}=\sqrt{kT_n/m_i}$ is the neutral sound speed at the boundary and $\gamma_n$ is a neutral wall heat transmission coefficient. Here we take $\gamma_n=0.25$ as a reasonable estimate derived from energy reflection coefficients for deuterium incident on a tungsten wall from the TRIM database\footnote{\url{http://www.eirene.de/html/surface_data.html}}.}

\subsubsection{Atomic state bundling}

The final change is to introduce bundling on the neutral states, so that a reduced number of states are evolved without significantly altering the plasma-neutral physics. This is desirable as it reduces the number of electron-neutral collision operators in (\ref{eq:vfp}) which need to be computed, which increases with the square of the number of evolved neutral states and therefore represents a code bottleneck. 

{We will define a cut-off neutral state index $\tilde{b}$, and} evolve all states from $b=1$ to $\tilde{b}-1$ as normal, and group all higher states (from $\tilde{b}$ to $b_{max}=30$) into a single bundle $\beta$, with bundle density $n_{\beta}=\sum_{b \in \beta} n_b$. 
For this bundle, we assume a Boltzmann state distribution at the electron temperature, giving the density ratio $n_{b}/n_{\tilde{b}}=(b / \tilde{b})^2 e^{-\left(\varepsilon_{\tilde{b}}-\varepsilon_{b}\right) / k T_{e}}$ for $b\ge \tilde{b}$, where $\varepsilon_b$ is the ionisation energy of level $b$, which for hydrogen is $\varepsilon_b=13.6\mathrm{eV}/b^2$. 

All particle, momentum and energy source terms relating to this bundle can then be computed by evaluating the appropriate collision operator with bundle-averaged cross-sections, $\langle \sigma \rangle_{\beta}$ and ionisation energies, $\langle \varepsilon \rangle_{\beta}$. These quantities are functions of the electron temperature, and are precomputed on a relevant range of $T_e$ and interpolated at runtime. For example, the electron energy sink from ionisation of neutrals in $\beta$ is 

\begin{equation}
    \begin{split}
        Q_{\beta}^{ion} &= \langle \varepsilon \rangle_{\beta}  S_{\beta}^{i o n} \\
        &= \langle \varepsilon \rangle_{\beta} n_{\beta}n_e \langle K\rangle_{\beta}^{ion} \\
        &= \langle \varepsilon \rangle_{\beta} n_{\beta} n_e 4 \pi \int_{\infty} d v v^{3} \frac{f_{0}(v)}{n_e}\langle\sigma\rangle_{\beta}^{ion},
    \end{split}
\end{equation}
where $S_{\beta}^{ion}$ is the ionisation particle source from $\beta$, $\langle K\rangle_{\beta}^{ion}$ is the bundle-averaged ionisation rate coefficient, $n_e$ is the electron density and $f_0$ is the isotropic part of the electron distribution. We also include an effective collision operator to capture excitation collisions within a bundle by assuming the net excitation rate is equal to the rate of radiative de-excitation. 

Using $\tilde{b}=5$, this bundling method is found to predict neutral densities and radiative losses to within a few percent of that predicted by directly evolving all states, while providing {around an order of magnitude} reduction in compute time. Further details of this bundling approach will be published separately.  

\section{Parameter scan simulations}
\label{sec:simulations}

\begin{figure}
    \centering
        \includegraphics[width=0.6\linewidth]{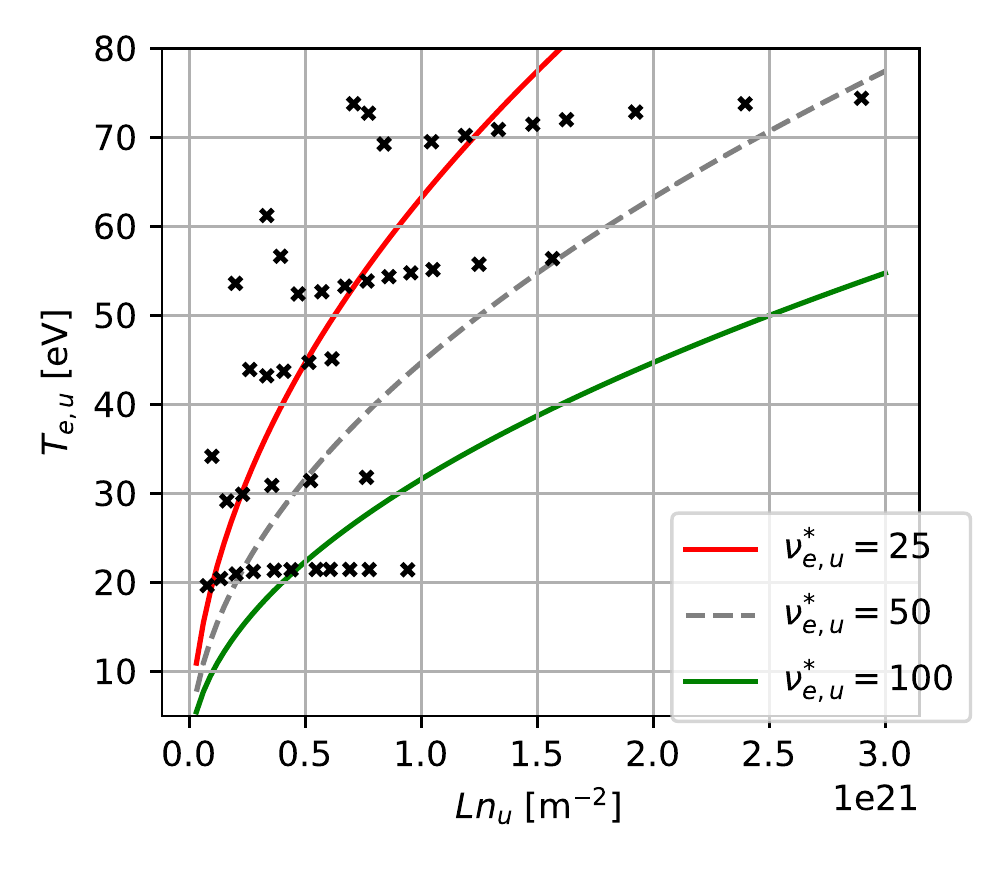}
        \caption{{SOL-KiT simulations carried out for this study, where each black dot represents a simulation at a given $q_{in}$ and $\langle n_0 \rangle$. The exact values of $n_u$ and $T_{e,u}$ are taken from the kinetic simulations. For context, lines of constant collisionality are shown at $\nu^*_{e,u}=$ 25, 50 and 100.}}
    \label{fig:sims}
\end{figure}

Two SOL parameters which we have some degree of control over in tokamaks (and which determine SOL behaviour) are $T_u$ and $n_u$, the plasma temperature and density measured at some upstream location. In these simulations, we vary these by changing the input power flux to the SOL from the core, $q_{in}$, and the initial line-averaged plasma density, {$\langle n_0 \rangle = \frac{1}{L} \int_0^L n_e(t=0) dx$}, where the plasma is fully ionised at initialisation. {The total number of particles (plasma plus neutrals) stays constant in the simulations due to 100\% recycling of the target particle flux.} {The input heat flux is distributed uniformly across the heating region, $L_{heat}$, as a volumetric energy source, $Q_{in}$, such that $q_{in}=\int_0^{L_{heat}}Q_{in}dx=L_{heat}Q_{in}$.}

Of interest in this study is how conditions upstream determine the electron transport, and a useful measure of this is the electron upstream collisionality parameter $\nu_{e,u}^*$, defined as the ratio of the connection length $L$ to the electron Coulomb mean free path upstream $\lambda_{ee,u}$ \cite{Stangeby2001}, 
\begin{equation}
    \nu_{e,u}^* = \frac{L}{\lambda_{ee,u}} \simeq 10^{-16} \frac{Ln_u}{T_{e,u}^2}
\end{equation}
for $T_{e,u}$ in [eV] and $n_u$ in [m$^{-3}$]. Note that this differs slightly from some forms of $\nu^*=L/\lambda_u$ employed in the literature (e.g. \cite{Wigram2020}), and $\nu_{e,u}^*$ here will typically be smaller than collisionality defined in terms of total plasma temperature because $T_{i,u}>T_{e,u}$ generally. 

{For a deuterium plasma, a number of density scans were performed at different input powers and connection lengths. Connection lengths ranged from $L=11.93$m to 30.97m, input powers from $q_{in}=4$MWm$^{-2}$ to 128MWm$^{-2}$, and densities from $\langle n_0 \rangle=1.0 \times 10^{19}$m$^{-3}$ to $1.4 \times 10^{20}$m$^{-3}$. With these input parameters, the simulations cover $\nu_{e,u}^*$ from 6.3 to 203.6. At the lowest collisionalities the plasma is sheath-limited, while detachment is reached at the highest values of $\nu^*_{e,u}$ (measured by rollover of the target particle flux).}

$q_{in}$ is distributed over {approximately the first third of the domain} and spread equally between the ions and electrons; 100\% of plasma particles lost to the sheath are recycled as neutrals, and the pitch angle used in the neutral model was $\alpha=15^{\circ}$. 100 spatial cells were used, which were spaced logarithmically with higher resolution close to the target. {For the simulations with the longest connection length, the spatial grid widths ranged from 2.28m upstream to 1.05mm at the target}. 
In velocity space (for kinetic electron runs), a geometric grid of 80 cells was used up to a velocity of $\simeq$ 12$v_{th,0}$,where $v_{th,0}$ is the thermal velocity of electrons at a reference temperature of 10eV. The resolution was higher at low velocities, such that grid widths ranged from 0.05$v_{th,0}$ to 0.35$v_{th,0}$. In the kinetic runs, the kinetic equation (\ref{eq:vfp}) was solved up to the spherical harmonic $l_{max} = 3$.

To reach equilibrium, determined by when the power and particle balance has converged, the kinetic simulations with SOL-KiT each take a few weeks running on 8 CPUs, while the fluid simulations typically take a day or less on 4 cores. 

These simulations are situated on the $T_{e,u}$-$Ln_u$ plane along with lines of constant $\nu_{e,u}^*$ in Figure \ref{fig:sims}. For reference, present-day tokamaks (JET, DIII-D, etc.) operate with $Ln_u\simeq10^{20}-10^{21}$m$^{-2}$ and $T_{e,u}\simeq20-60$eV. {Future devices like ITER and DEMO will operate with $Ln_u\sim1\times 10^{21}- 4 \times 10^{21}$m$^{-2}$ and $T_{e,u}\sim150-300$eV\mbox{\cite{Veselova2021,Rubino2017}}}. Simulating such regimes kinetically is computationally demanding, {but this study represents an attempt to explore kinetic effects in regimes beyond those attainable in existing devices. The simulations presented here fall short of reaching the highest values of $T_{e,u}$ expected in future tokamaks, but do encompass reactor-relevant values of $Ln_u$, detached conditions, and a broad range of upstream collisionalities.} 

\section{Results}
\label{sec:results}

\begin{figure}[h]
    \begin{subfigure}{.49\textwidth}
      \centering
      \includegraphics[width=\linewidth]{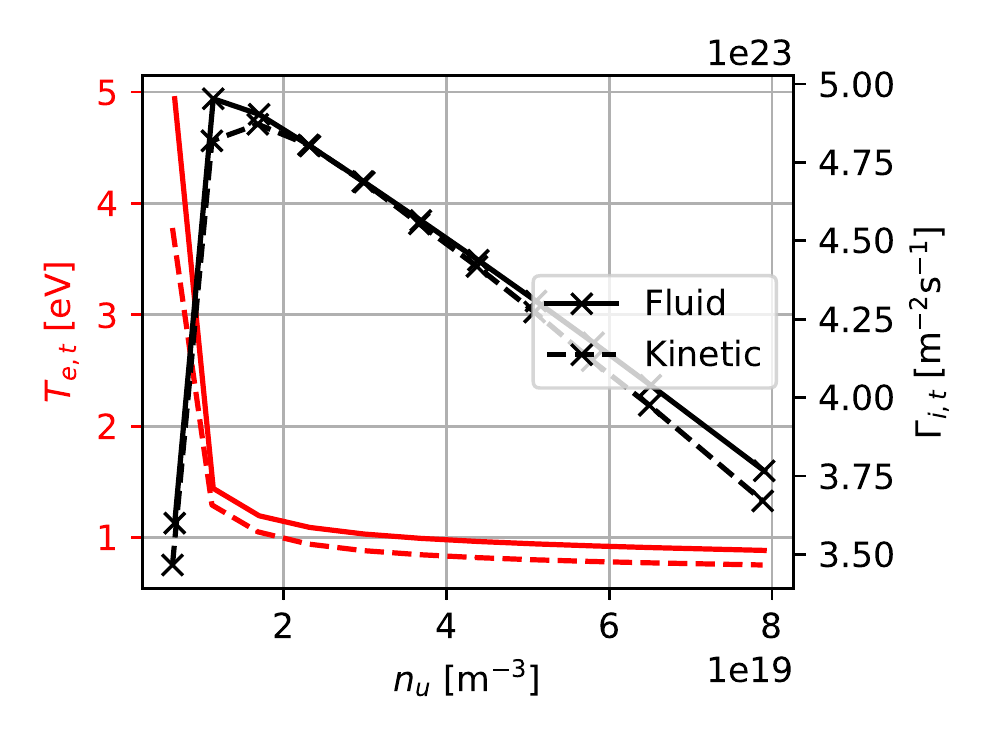}
      \caption{{$q_{in}=4$MWm$^{-2}$, $L=11.93$m.}}
      \label{fig:rollover_plots_p4}
    \end{subfigure}
    \hfill
    \begin{subfigure}{.49\textwidth}
      \centering
      \includegraphics[width=\linewidth]{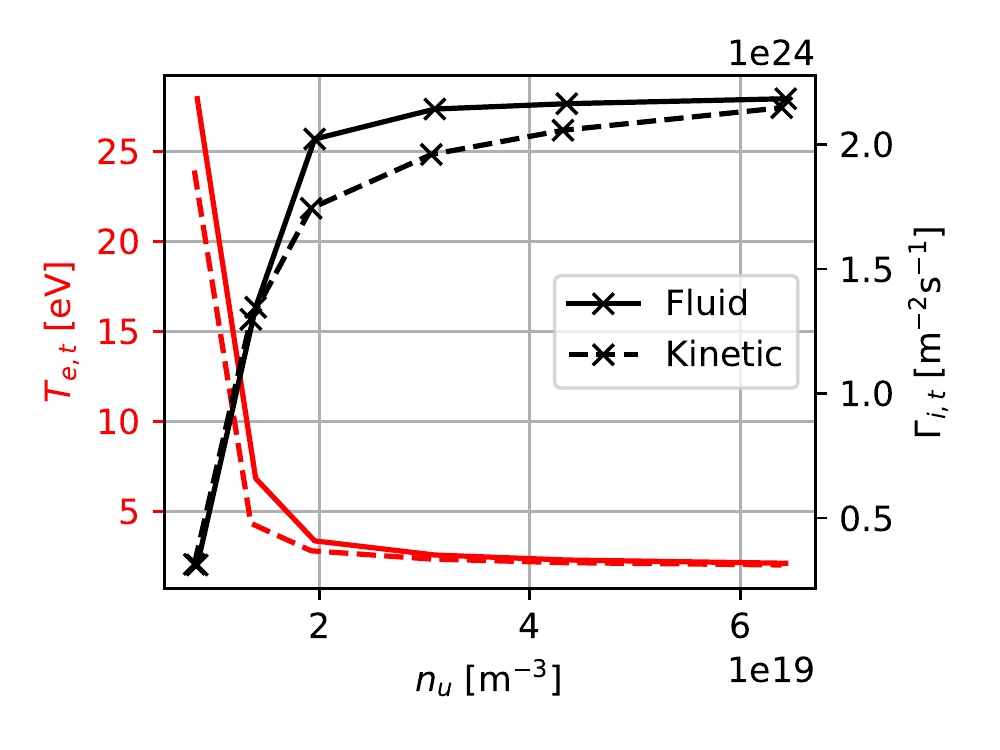}
      \caption{{$q_{in}=16$MWm$^{-2}$, $L=11.93$m.}}
    \end{subfigure}
    
    \begin{subfigure}{.49\textwidth}
      \centering
      \includegraphics[width=\linewidth]{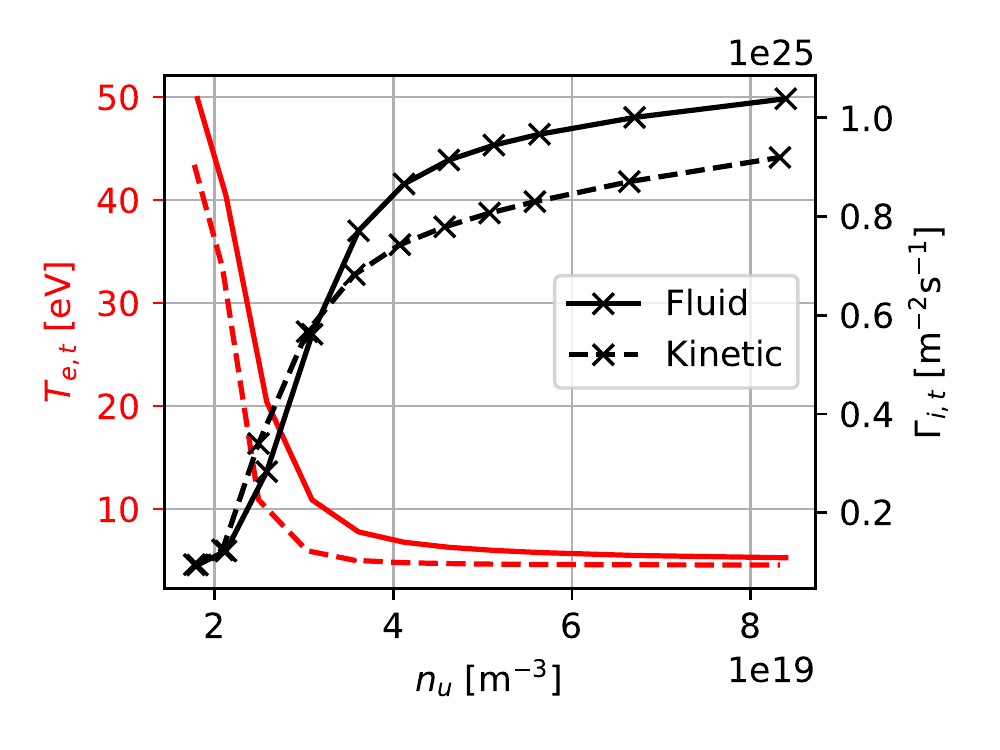}
      \caption{{$q_{in}=80$MWm$^{-2}$, $L=18.79$m.}}
    \end{subfigure}
    \hfill
    \begin{subfigure}{.49\textwidth}
      \centering
      \includegraphics[width=\linewidth]{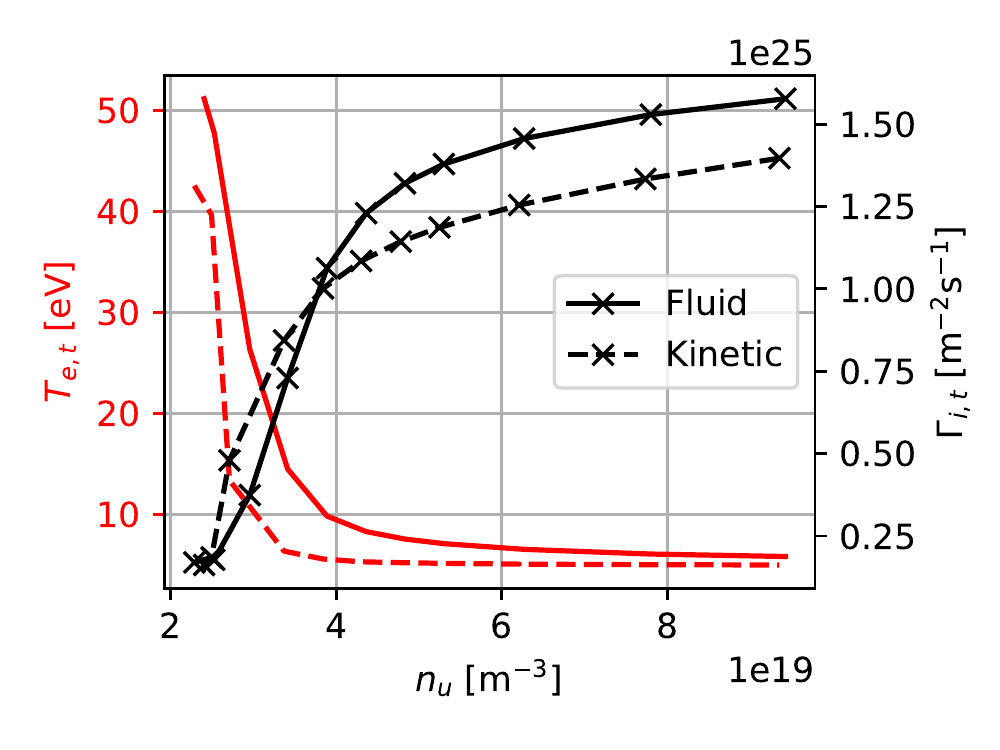}
      \caption{{$q_{in}=128$MWm$^{-2}$, $L=30.96$m.}}
    \end{subfigure}
    \caption{{Target electron temperatures, $T_{e,t}$, and ion fluxes, $\Gamma_{i,t}$, for four of the density scans at different input powers and connection lengths. Results with fluid and kinetic electrons are shown. The lowest input power runs exhibit detachment, indicated by rollover of $\Gamma_{i,t}$.}}
    \label{fig:rollover_plots}
  \end{figure}

We start by displaying in Figure \mbox{\ref{fig:rollover_plots}} the target temperatures and particle fluxes of {four of the density scans carried out.} Rollover of the target flux, an indicator of detachment onset, is expected when particle, momentum and power losses are sufficient to reduce the target flux despite increasing plasma density. Only the lowest-power run reaches flux rollover {here, while all other runs are partially or fully attached.} 
{The absence of rollover at high input powers is not unexpected given the lack of impurity radiation in the SOL-KiT model, which would provide additional power dissipation.} 
{It can be seen in Figure \mbox{\ref{fig:rollover_plots}} that target temperatures are lower when the electrons are treated kinetically, with the biggest differences in both absolute and relative terms occurring at low upstream densities. However, there are only small differences observed in the rollover behaviour in the one density scan which does reach detachment, \mbox{\ref{fig:rollover_plots_p4}}, where a kinetic treatment of the electrons does not change the position of detachment onset when varying $n_u$, i.e. the qualitative behaviour of $\Gamma_{i,t}$ is the same. There is a decrease in $\Gamma_{i,t}$ with kinetic electrons at high upstream densities, up to around 15\%, while there is a slight increase at low densities. }

\begin{figure}[t]
    \begin{subfigure}[t]{.47\textwidth}
      \includegraphics[width=\linewidth]{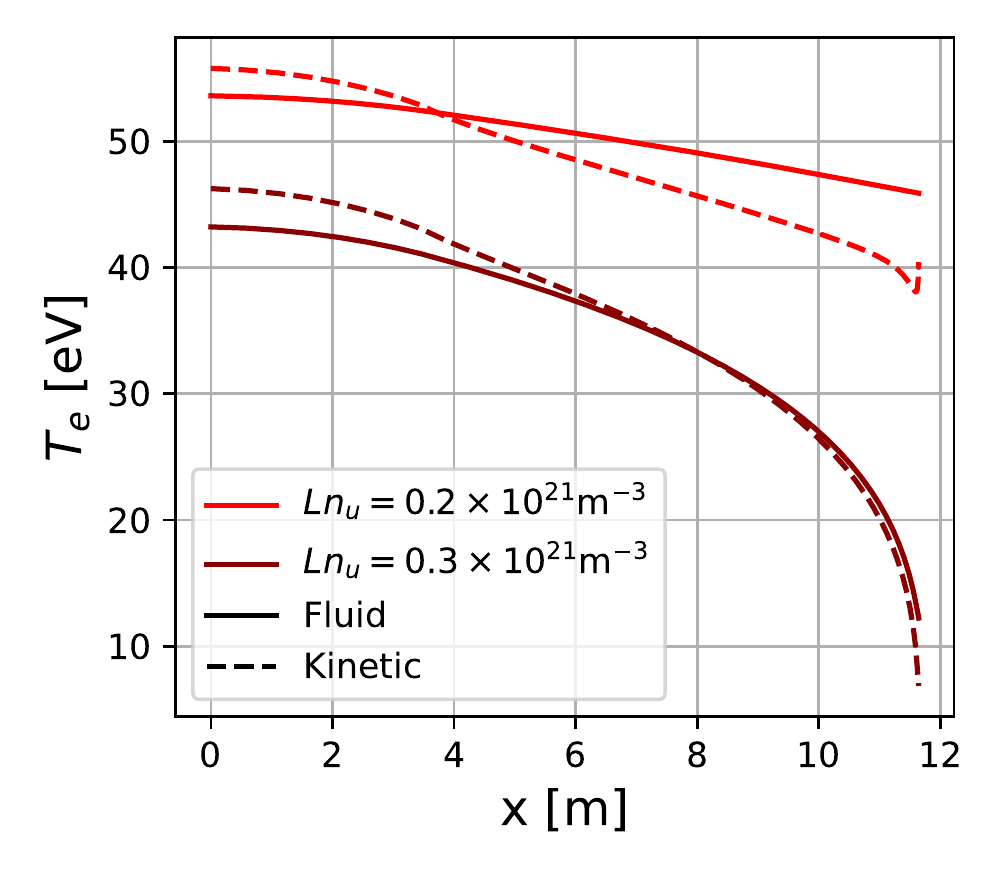}
      \caption{Fluid and kinetic electron temperature profiles.}
      \label{fig:flux_suppressiona}
    \end{subfigure}
    \begin{subfigure}[t]{.47\textwidth}
      \includegraphics[width=\linewidth]{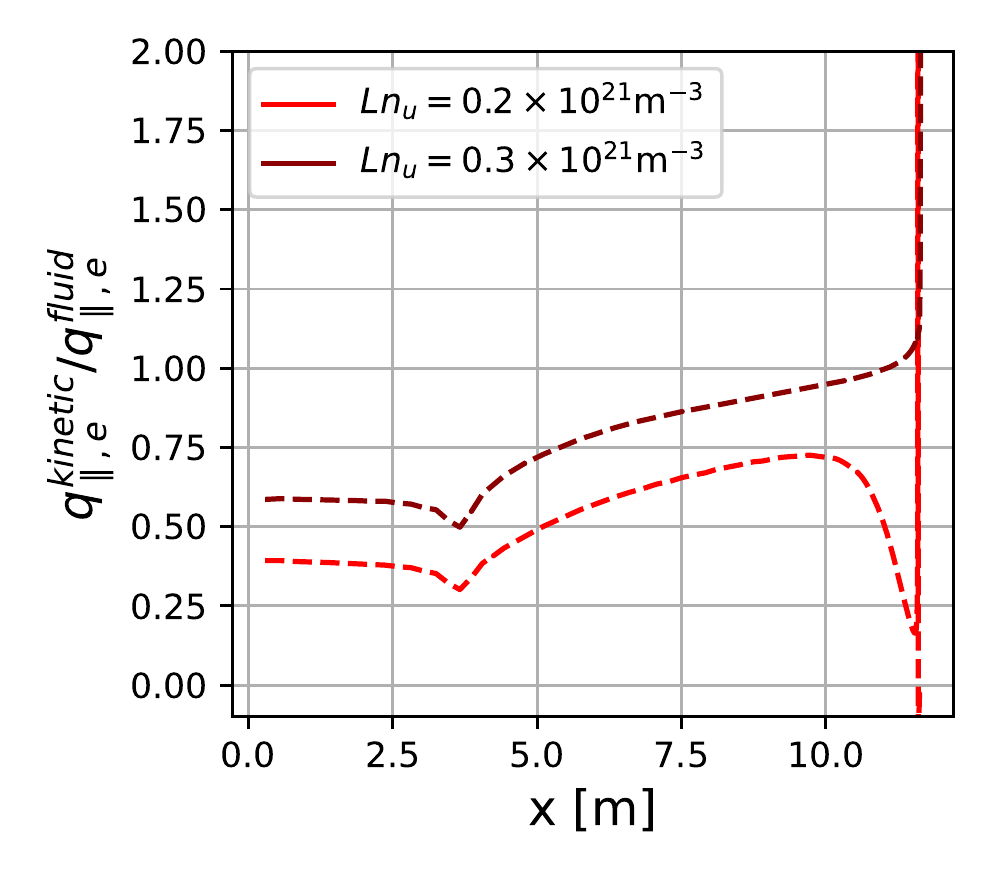}
      \caption{Ratio of the kinetic to Spitzer-H\"{a}rm conductive heat flux for the kinetic simulations in (a).}
      \label{fig:flux_suppressionc}
    \end{subfigure}
    \caption{{Kinetic heat flux suppression resulting in steeper temperature gradients and lower target temperatures for two low collisionality simulations ($q_{in}=64$MWm$^{-2}$, $L=11.93$m).}}
    \label{fig:flux_suppression}
  \end{figure}

\begin{figure}[t!]
  \centering 
  \includegraphics[width=0.7\textwidth]{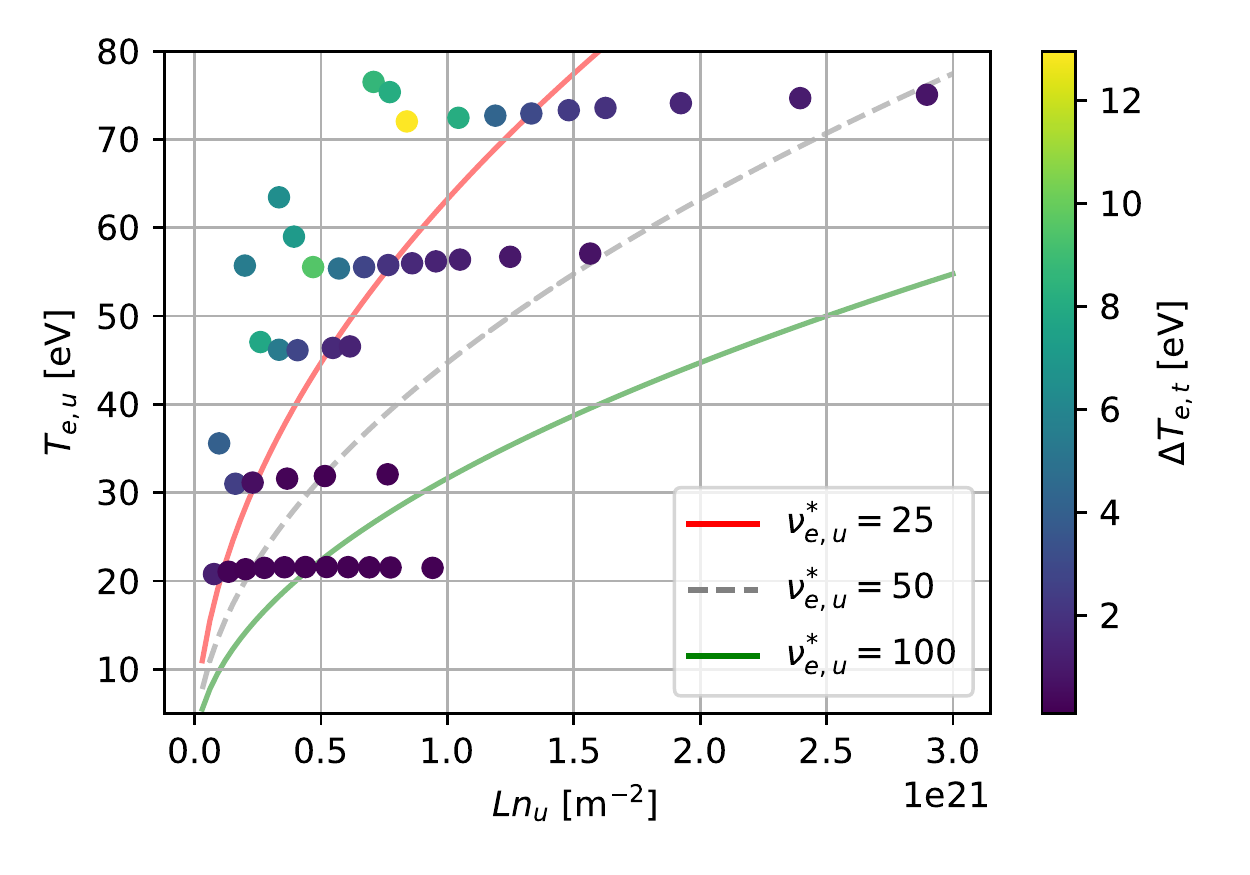}
  \caption{{Difference in target temperatures for kinetic vs. fluid electrons across all simulations, $\Delta T_{e,t} = T_{e,t}^{fluid} - T_{e,t}^{kinetic}$.}}
  \label{fig:T_t_diff}
\end{figure}

The reduction in target temperatures in Figure \ref{fig:rollover_plots} is a reflection of {the suppression of the parallel conductive electron heat flux}
which is observed in kinetic simulations, as has been observed in other kinetic studies of parallel transport \cite{Chankin2018,Mijin2019,Wigram2019,Brodrick2017}. This can be seen in Figure \ref{fig:flux_suppressiona}, which shows temperature profiles for two simulations at low collisionality, along with differences in target electron temperatures across all simulations in Figure \ref{fig:T_t_diff}. Figure \ref{fig:flux_suppressionc} shows the ratio of the kinetic to Spitzer-H\"{a}rm heat flux calculated on the kinetic plasma profiles in \ref{fig:flux_suppressiona}. This suppression of the heat flux relative to that predicted by a fluid treatment, where for a given heat conductivity $\kappa$ the heat flux is $q_{\parallel,e} = -\kappa \nabla T$, arises due to fast electrons not depositing their energy locally due to their large mean free path relative to the temperature gradient length scale. This means that a steeper temperature gradient is required to achieve the same heat flux along the SOL, which is fixed by $q_{in}$. {There is an uptick in $T_e$ close to the wall in the least collisional simulations, which is seen in the hotter $T_e$ profile in Figure \mbox{\ref{fig:flux_suppressiona}} and as a spike in the heat flux ratio in Figure \mbox{\ref{fig:flux_suppressionc}}.}

\begin{figure}[h]
    \centering
        \includegraphics[width=0.6\linewidth]{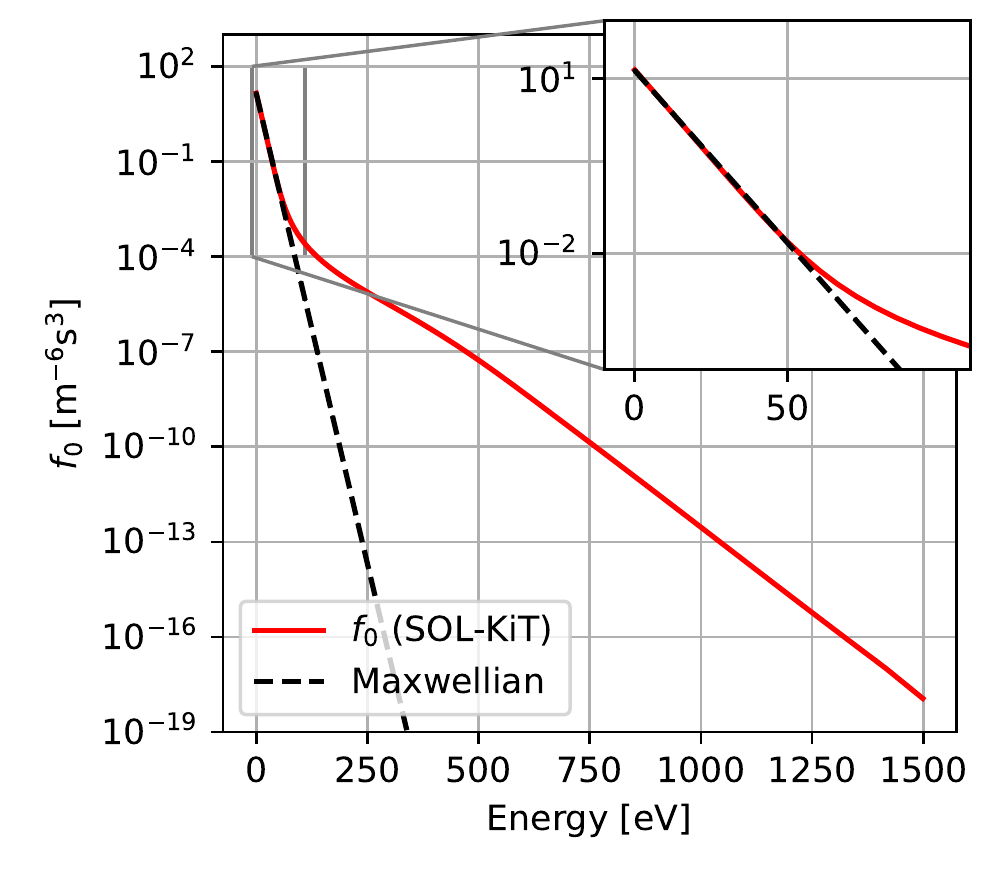}
        \caption{Electron energy distribution (isotropic part) close to the wall in a SOL-KiT simulation ($q_{in}=64$MWm$^{-2}$, $L=11.93m$, $\langle n_0 \rangle =5\times10^{19}$m$^{-3}$). Dashed line is the local Maxwellian. A prominent high-energy tail and thermalised bulk can both be seen.  $T_e=7.3$eV, $n_e=3.2\times 10^{20}$m$^{-3}$.}
    \label{fig:f0}
\end{figure}

{In Figure \mbox{\ref{fig:f0}} we can see the accumulation of fast electrons from upstream at the target, where an electron energy distribution from just in front of the sheath boundary is shown.} There is a clear enhanced high-energy tail, while the thermal bulk is close to the local Maxwellian.

\begin{figure}[h]
  \centering
  \includegraphics[width=0.7\textwidth]{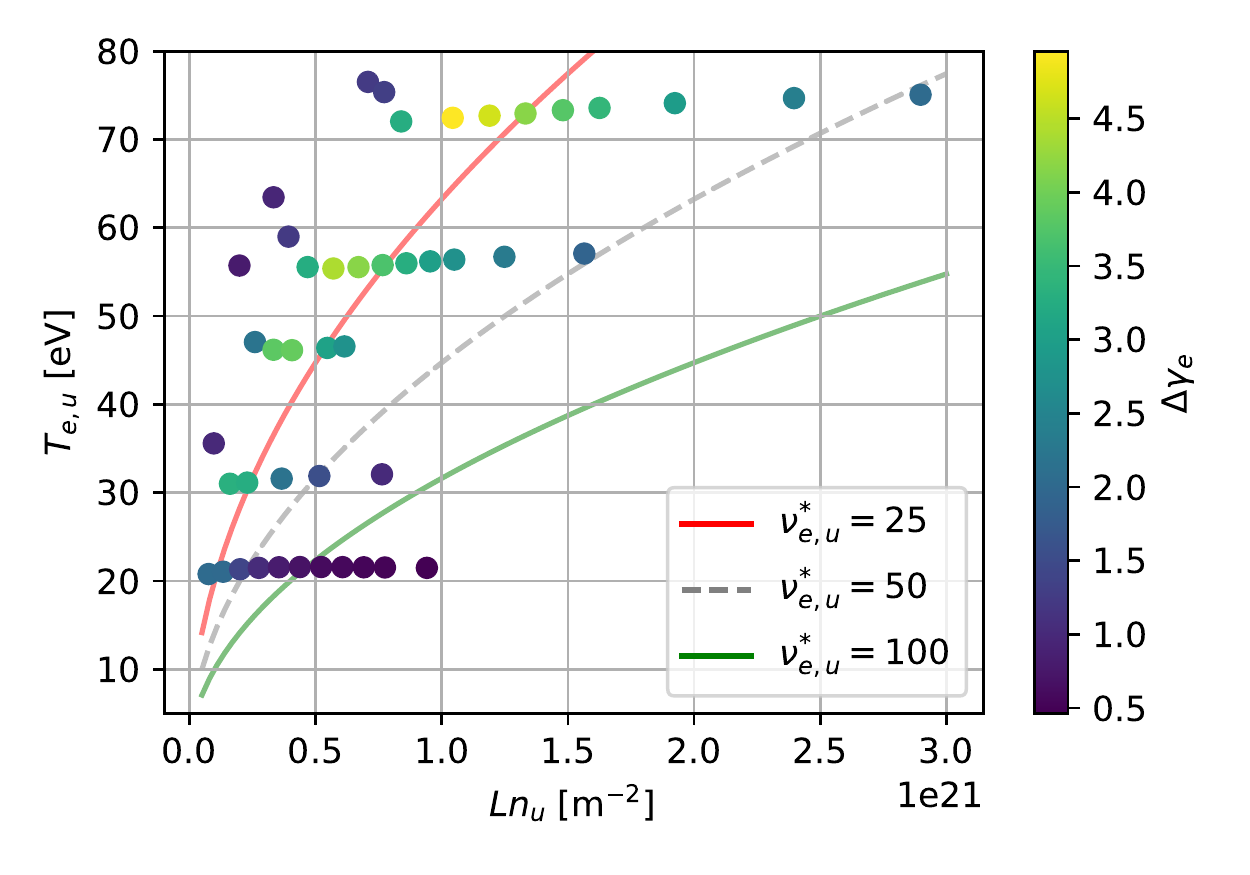}
  \caption{{Enhancement to the electron sheath heat transmission coefficient, $\Delta \gamma_e=\gamma_e^{kinetic} - \gamma_e^{fluid}$, across all simulations.}}
  \label{fig:gamma_e_enhancement}
\end{figure}

In Figure \ref{fig:gamma_e_enhancement}, we show the kinetic enhancement of the sheath heat transmission coefficient, $\gamma_e$, {shown as $\Delta \gamma_e=\gamma_e^{kinetic} - \gamma_e^{fluid}$. Maximum and minimum values of $\Delta \gamma_e$ seen here are 4.96 and 0.47.}
Differences in $\gamma_e$ here arise because, in kinetic mode, $\gamma_e$ is calculated self-consistently from the logical boundary condition on the electron distribution, whereas in fluid mode it is calculated from fluid quantities in the classical way (\ref{eq:gamma_e_def}). {For reference, in fluid mode typically $\gamma_e\simeq 4.8$, and this varies slowly with SOL conditions.}
{In Figure \mbox{\ref{fig:gamma_e_enhancement}}, there is a non-monotonic behaviour in $\Delta \gamma_e$, where the classical value is approached at both high and low collisionalities.}
Similar behaviour was seen in a power scan in \cite{Mijin2019} and in a collisionality scan in \cite{Zhao2017}. The largest differences occur at low-intermediate collisionalities, but there is an additional increase in magnitude of this effect along lines of constant collisionality, moving towards larger $T_{e,u}$ and $n_u$. This can be seen by tracing along the red $\nu^*_{e,u}=25$ line in Figure \ref{fig:gamma_e_enhancement}, where simulations at higher $T_{e,u}$ have larger $\Delta \gamma_e$. Additionally, even at the highest collisionalities, where we would expect good agreement between fluid and kinetic predictions, there is a residual $\Delta \gamma_e \simeq 0.5$. It would therefore appear that convergence of $\gamma_e$ to the fluid value is slow as a function of $\nu_{e,u}^*$. 

\begin{figure}[h]
    \centering
        \includegraphics[width=0.8\linewidth]{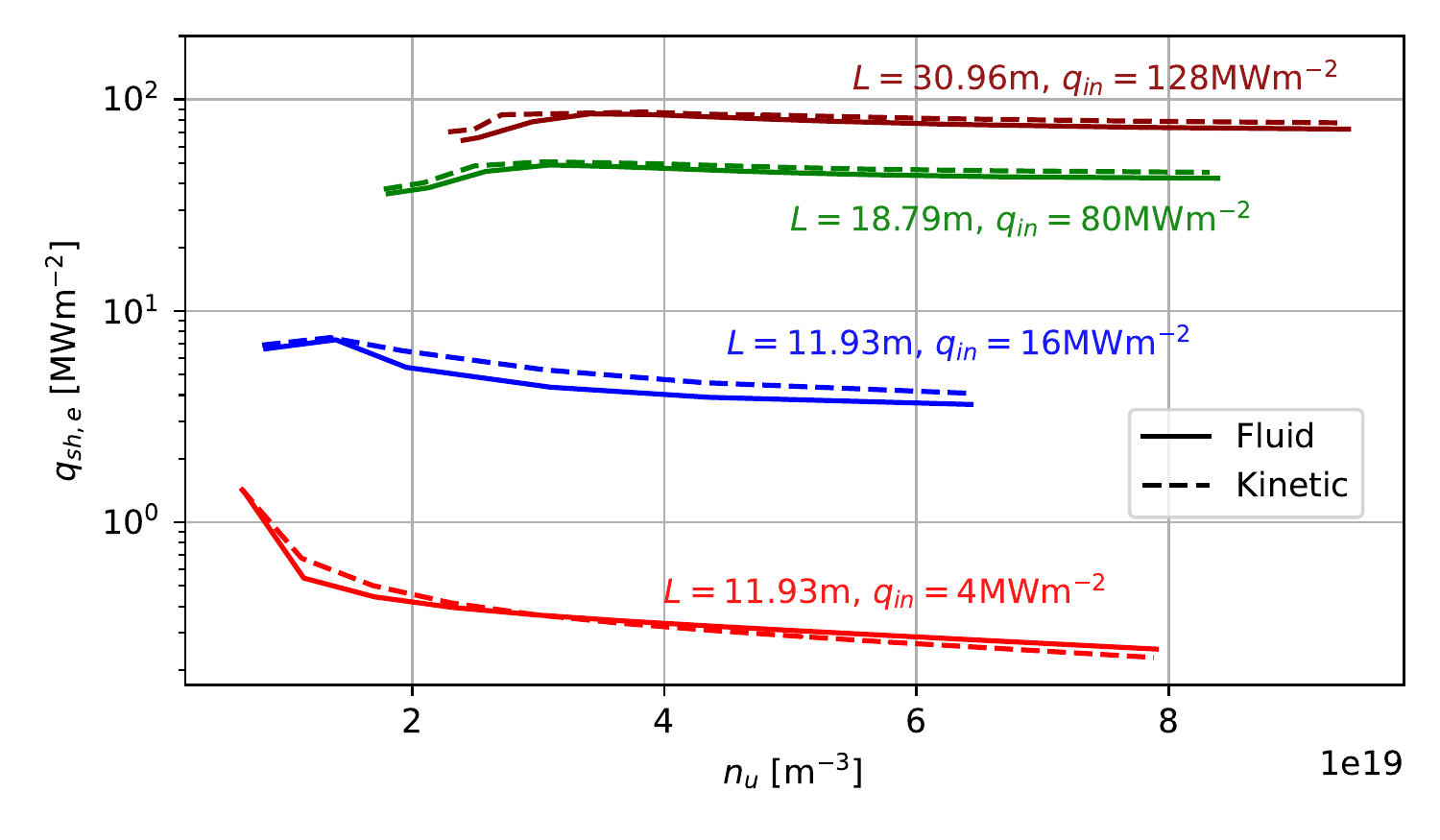}
        \caption{{Sheath heat flux from electrons, $q_{sh,e}$, for kinetic and fluid simulations, grouped into density scans at different connection lengths and input powers.}}
    \label{fig:q_sh}
\end{figure}

Given the enhancement in $\gamma_e$ for kinetic electrons, it is natural to investigate the heat lost to the sheath boundary, $q_{sh,e}=\gamma_e kT_{e,t} \Gamma_{i,t}$, where $T_{e,t}$ is the electron temperature at the target. This is shown in Figure \ref{fig:q_sh}{, where the variation of $q_{sh,e}$ for kinetic and fluid simulations with $n_u$ is plotted for simulations grouped by connection length and input power.}. In contrast to the kinetic enhancement in $\gamma_e$, we see that $q_{sh,e}$ is generally in good agreement for kinetic and fluid simulations. This is perhaps not surprising, since $q_{sh,e}$ is to a large extent fixed by $q_{in}$, as well as the fact that kinetic enhancement in $\gamma_e$ may be offset by the reduction in target temperatures (Figure \ref{fig:flux_suppressiona}). However, this does show that the overall power balance in these simulations (for example, how much power is radiated away by electron-neutral collisions) is broadly unchanged despite modifications to the conductive transport as well as behaviour at the boundary. 

Finally, it is worth commenting that the findings in \cite{Mijin2019} and \cite{Zhao2019}, that electron-neutral reaction rates are well-approximated by Maxwellian-averaged rates, is replicated in this study. Differences in rate coefficients for deuterium ionisation and line radiation are negligible. Some differences in the total line-integrated particle source do exist in kinetic simulations (which takes into account ionisation and recombination as well as multi-step processes involving ex/de-excitation), but these are all under 10\%, and are driven purely by differences in temperature profiles.

\section{Discussion}
\label{sec:discussion}


{The detachment behaviour observed in these simulations contrasts somewhat with the observation of flux rollover at $n_u \sim 2 \times 10^{19}$m$^{-3}$ in the study by Dudson et al. in\mbox{\cite{Dudson2019}}, which uses a similar simulation setup for the SD1D fluid code with $q_{in}=50$ MWm$^{-2}$ and $L=30$m. With the model presented here, $q_{in}$ must be reduced to below 25 MWm$^{-2}$ to reach detachment at this connection length. A separate investigation has highlighted that an underestimation of the ionisation rate at high electron densities in SD1D may be the part of the reason for this difference\mbox{\cite{Kryjak2022}}, as well as the fact that SOL-KiT does not include impurity radiation or flux tube expansion.}

{The uptick in $T_e$ close to the target in the least collisional simulations is seen in the hotter $T_e$ profile in Figure \mbox{\ref{fig:flux_suppressiona}} and in a spike in the heat flux ratio in Figure \mbox{\ref{fig:flux_suppressionc}}, where $q_{\parallel,e}^{kinetic}$ can become negative in some simulations. This is the reason why the least collisional simulations do not have the largest differences in target temperatures (Figure \mbox{\ref{fig:T_t_diff}}) despite exhibiting higher flux suppression, where the uptick in $T_e$ close to the wall to some extent cancels the increased temperature gradient. This feature comes primarily from the perpendicular electron temperature, which has been seen elsewhere \mbox{\cite{Tang2015,Zhang2022a}}, where it is proposed to be related to an enhancement in the parallel flux of the perpendicular electron energy.}

{The results in Figure \mbox{\ref{fig:rollover_plots}} suggest that a kinetic treatment of electron transport does not predict easier access to detachment. However, it should be said that only the most collisional density scan considered here actually reaches detachment, which is where kinetic effects are weakest, and it is possible that differences in detachment thresholds may exist in conditions with higher $T_{e,u}$. }

The largely unchanged target flux behaviour in simulations with kinetic electrons, along with broadly similar heat loads to the walls (Figure \ref{fig:q_sh}), is indicative of the fact that a kinetic electron treatment does not significantly change the particle, momentum or power balance at equilibrium in this 1D SOL model. This is despite strong heat flux suppression (Figure \ref{fig:flux_suppression}) and enhancement of the sheath heat transmission coefficient (Figure \ref{fig:gamma_e_enhancement}). This can be understood as resulting from the fact that heat transport is primarily determined by the input power $q_{in}$. While a modified temperature profile is needed in kinetic mode to achieve the same parallel heat flux in these simulations, this is compensated by an enhanced $\gamma_e$ which gives a similar $q_{sh,e}$, leaving the power balance broadly unchanged. In addition, differences in the temperature profile are insufficient to significantly change the particle source from electron-neutral interactions. 

This power balance behaviour would not necessarily continue to be the case in the presence of strong radiation sinks from impurities, where modified temperature profiles and reaction rates could lead to differences in overall energy transport. This is the subject of an ongoing study.

The unchanged power balance despite the presence of kinetic effects in parallel heat flux and $\gamma_e$ suggests that any attempt to capture kinetic effects in a fluid framework would need to consider both phenomena. As such, approaches which treat only the modified heat flux \cite{Wigram2020} or the boundary condition \cite{Vasileska2020} may not provide better predictive power than a purely fluid model. 

The strong enhancements to $\gamma_e$ are a result of the modified potential drop across the sheath when calculated kinetically. {This depends on $v_c$, the cut-off velocity at which the electron distribution at the sheath is truncated. The value of $v_c$ is set to ensure ambipolar particle flux, and is therefore somewhat sensitive to a strongly enhanced tail of the electron distribution, of the kind observed in Figure \mbox{\ref{fig:f0}}. This may therefore have consequences for measurements of the electron heat flux, which requires knowledge of $\gamma_e$, and the electron temperature, where electrons are assumed to be close to Maxwellian. In \mbox{\cite{Tskhakaya2011}}, Tskhakaya et al. observed large discrepancies in simulated Langmuir probe measurements of $T_e$ due to the departure of the electrons from Maxwellian. Similarly to the study presented here, the largest discrepancies were seen at intermediate collisionalities.}

{It is worth noting that the most collisional kinetic simulations here exhibit a $\gamma_e$ which is $\sim0.5$ higher than the classical value (\mbox{\ref{eq:gamma_e_def}}). Since the electron and ion momenta are treated separately and used to solve for the electric field in this model, this residual discrepancy is not from the pre-sheath acceleration of the ions. Instead, it is a result of the departure of the electrons from Maxwellian at the wall.}

As discussed, and shown in \cite{Mijin2019} and \cite{Zhao2019}, electron-impact ionisation rates of hydrogen are very well approximated by a Maxwellian distribution. This is unsurprising when considering the distribution shown in Figure \ref{fig:f0}, which is non-Maxwellian in the tail but very close to Maxwellian in the thermal bulk. Given the energy threshold of inelastic processes involving hydrogen are all at or lower than 13.6eV, the Maxwellian bulk electrons dominate the rates. This does suggest however that inelastic processes with threshold energies $\gtrsim$ 50eV (e.g. ionisation of high-Z impurities) may exhibit strong kinetic enhancement due to the presence of this enhanced tail. An ongoing study is currently investigating this. 


\begin{figure}
    \centering
        \includegraphics[width=0.5\linewidth]{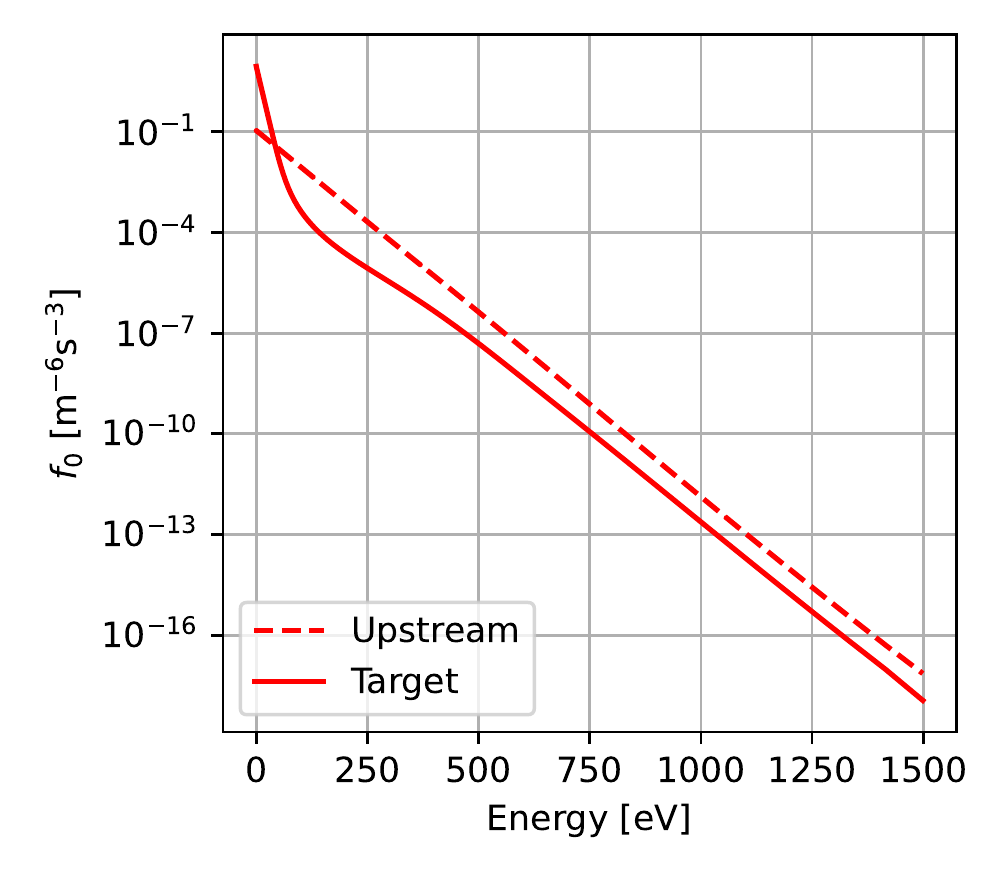}
        \caption{Electron energy distributions at two locations, upstream ($T_e=39.9eV$) and close to the wall ($T_e=8.0$eV). Electrons are close to Maxwellian upstream, and the fast tail survives to some extent further downstream. {The gradient of the tail, related directly to the temperature on these axes, is the same in both cases.}}
    \label{fig:imprint}
\end{figure}

{The enhanced tail of the electron distribution seen close to the wall in these simulations is an imprint from upstream, as can be seen clearly in}
Figure \ref{fig:imprint}, where the upstream distribution is plotted alongside that close to the target. 
Therefore, if we assume the tail of the distribution at the target has temperature $T_{e,t}^{tail}=T_{e,u}$, then two conditions existing simultaneously produce a `strongly enhanced' tail, which can lead to strong kinetic effects as discussed. {These are $T_{e,u}\gg T_{e,t}$ and $\nu^*_{e,u}$}. 
In this study, we see that the imprint can survive up to moderate values of $\nu^*_{e,u}$ and hence drive kinetic effects, for example in the peak enhancement to $\gamma_e$ occurring at $\nu^*_{e,u}\simeq20$ (Figure \ref{fig:gamma_e_enhancement}). It is the interplay of upstream collisionality and parallel temperature drops which determines the strength of this imprint. For tokamak edge plasmas with large $T_{e,u}$ as well as significant power dissipation via impurities, we might expect both of these conditions to be satisfied.

Contrary to the heat flux suppression, which appears to be a monotonic function of $\nu^*_{e,u}$,
the enhancement to $\gamma_e$ is more complex. It peaks at $\nu^*_{e,u}\simeq20$, but also appears to increase for increasing $T_{e,u}$ at constant $\nu^*_{e,u}$. If this behaviour can be extrapolated to reactor-class devices then we may expect significant deviations from classical values of $\gamma_e$. This is discussed further in the next section. 

\section{Scaling relationships for observed kinetic effects}
\label{sec:predictions}


Any attempt at capturing kinetic effects at equilibrium in a fluid model of the scrape-off layer would appear to need to capture both modifications to the heat flux and enhancement to the sheath heat transmission coefficient. While models do exist for the former \cite{DelSorbo2015,Schurtz2000,Ji2009}, they do not typically provide a self-consistent method for calculating modifications to the boundary behaviour. In \cite{Tskhakaya2008a}, Tskhakaya et al. provide fits to the modifications to $\gamma_e$ and parallel heat flux for the time-dependent response to a simulated edge-localised mode (ELM). Here, we present fits to the kinetic modifications to $\gamma_e$ and $q_{\parallel,e}$ seen at equilibrium across a range of $T_{e,u}$ and $n_u$ (and hence $\nu^*_{e,u}$), presented as functions of basic SOL parameters. 

{The approach taken in developing the relationships presented here has been to parameterise each simulation in terms of SOL quantities which are either control parameters or are easily obtainable from experiment or simulation. We then quantify the two strongest kinetic effects observed in the kinetic simulations, which we will call kinetic factors. These are
the line-averaged heat flux suppression}
\begin{equation*}
  f_{\kappa_e} = \frac{1}{L}\int q_{\parallel,e}^{kinetic}/q_{\parallel,e}^{SH} dx,
\end{equation*}
{and the enhancement to $\gamma_e$,}
\begin{equation*}
  \Delta \gamma_e = \gamma_e^{kinetic} - \gamma_e^{fluid}.
\end{equation*}
{We have then used a least squares fit, allowing a set of fit parameters to vary, to find combinations of these SOL parameters which provide the best predictive power for the kinetic factors. In general, it has been found that there are several forms of the relationships between these kinetic factors and SOL parameters which provide a comparably good fit. In this case, we have opted to present fits which are functions of a small number of SOL parameters to avoid over-fitting the data and keep the approach simple, and which also have physically realistic asymptotic behaviour (for example, $f_{\kappa_e}$ should be close to 1 at high collisionality). 

The fits presented here are in terms of control/upstream SOL parameters only. If there is knowledge of the target conditions, for example in a SOL fluid code, relationships with very good predictive power have been found, in particular for $\Delta \gamma_e$. One example is provided below. However, it has been found that some of these do not extrapolate well to conditions where the upstream collisionality is low and power dissipation is large, for example due to impurity radiation. Conditions of this nature have not been simulated yet with SOL-KiT, although this is planned for future work.
} 
\begin{figure}[t]
  \centering 
  \includegraphics[width=0.47\textwidth]{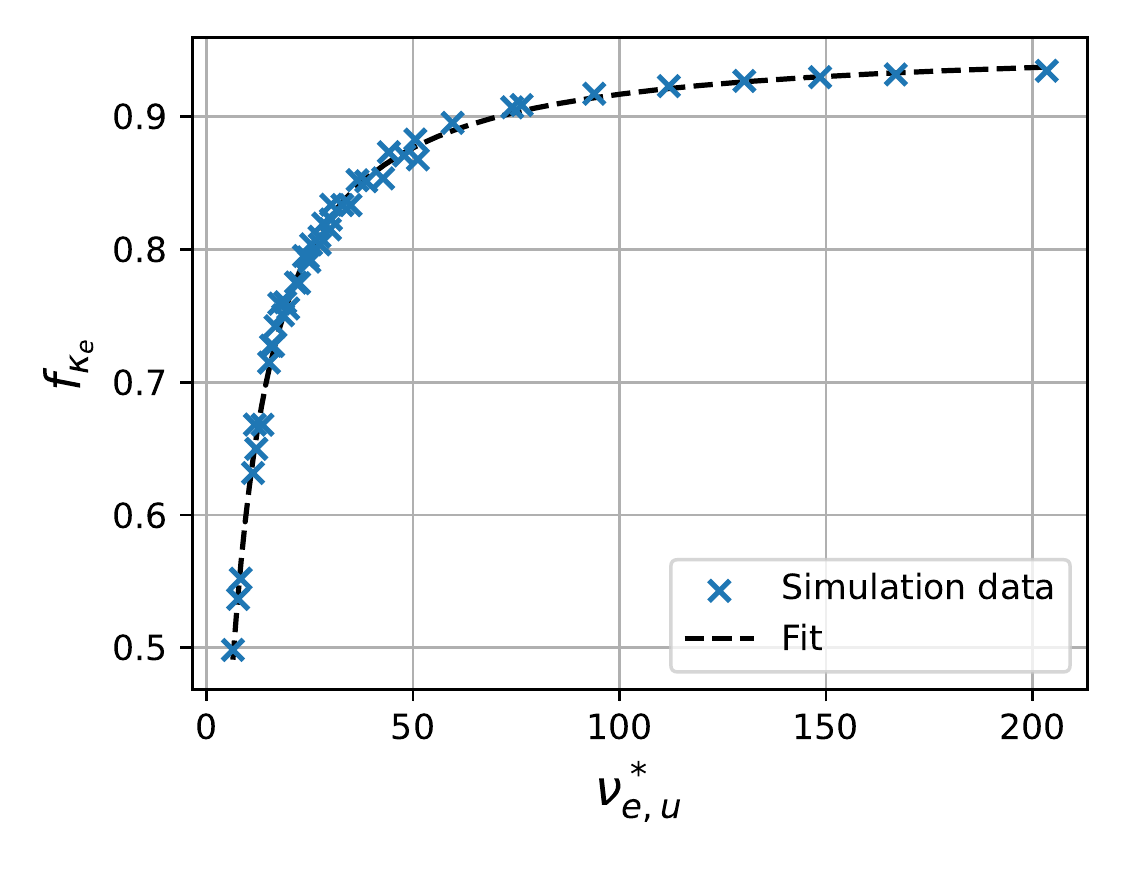}
  \caption{{The variation of $f_{\kappa_e}$ as a function of $\nu_{e,u}^*$ in simulations alongside the fit in equation (\mbox{\ref{eq:f_kappa_fit}}).}}
  \label{fig:f_kappa_fit}
\end{figure}

{The proposed relationship for the heat flux suppression is}
\begin{equation}
  f_{\kappa_e} = a  \exp{ (b (\nu_{e,u}^*)^{c})} + d,
  \label{eq:f_kappa_fit}
\end{equation}
{with}
\begin{equation*}
  a =0.696,\quad b = -8.059,\quad c = -1.074,\quad d = 0.260.
\end{equation*}
{This is a function of $\nu_{e,u}^*$ only, which was found to be sufficient for an accurate prediction of $f_{\kappa_e}$.} This is plotted against the simulation data in Figure \mbox{\ref{fig:f_kappa_fit}}, {where the RMS error on the fit is 0.01. There is clearly significant heat flux suppression at low upstream collisionalities. If this behaviour can be extrapolated beyond the region of $\nu_{e,u}^*$ explored here, this will lead to increased temperature gradients in sheath-limited regimes where the plasma would otherwise be expected to be nearly isothermal. This can be seen to some extent in the temperature profiles of the low collisionality runs in Figure \mbox{\ref{fig:flux_suppressiona}}.}

\begin{figure*}[t]
  \centering
    \includegraphics[width=0.5\linewidth]{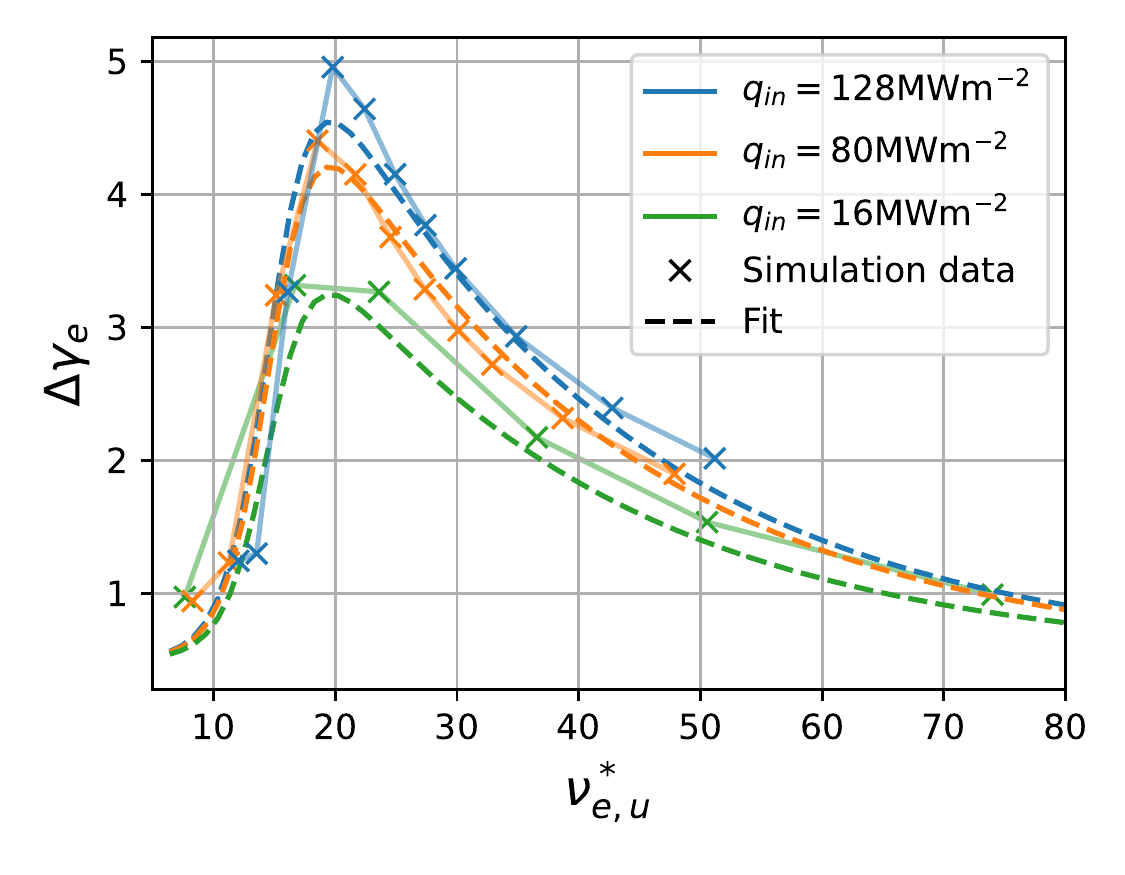}
    \caption{{Fit to $\Delta \gamma_e$ from equation (\mbox{\ref{eq:gamma_e_fit}}) as a function of $\nu_{e,u}^*$ for several values of $q_{in}$. Data from simulations with the same $q_{in}$ is shown alongside.}}
  \label{fig:gamma_e_fit}
\end{figure*}

{For the enhancement to the electron sheath heat transmission coefficient, we have}
\begin{equation}
  \Delta \gamma_e =  \frac{ a (q_{in})^{b} \exp(c \nu_{e,u}^*) }{ 1 + d \exp(e \nu_{e,u}^*) } + 0.5
  \label{eq:gamma_e_fit}
\end{equation}
with 
\begin{equation*}
  a = 9.93 \times 10 ^{-4},\quad b = 0.186 ,\quad c = 0.514 ,\quad d = 2.62 \times 10^{-4}, e = 0.553.
\end{equation*}
{This is shown in Figure \mbox{\ref{fig:gamma_e_fit}} and compared to simulation data for several values of $q_{in}$. The RMS error on this fit across all simulation data is 0.18. It can be seen that this fit captures both the peak to $\Delta \gamma_e$ at $\nu_{e,u}^* \sim 20$ and that fact that the peak increases slowly with $q_{in}$. It also drops to 0.5 for $\nu_{e,u}^* \lesssim 7$, but there is some disagreement with the simulation data in this regime so precise values may be higher. This low collisionality behaviour is somewhat speculative, but it is reasonable to expect that $\Delta \nu_{e,u}^*$ will at any rate be small at low collisionalities. This is because temperature gradients, which appear to be a necessary condition for significant enhancement to $\gamma_e$, are likely to be smaller at very low collisionalities than for intermediate $\nu_{e,u}^*$. This is, however, complicated somewhat by the increased heat flux suppression predicted at low collisionalities.
}

{By including conditions at the target into the fits, and considering relationships of the form $\Delta \gamma_e = a_0 \prod_i X_i^{a_i}$, where $a_i$ are fit parameters and $X_i$ are SOL parameters, it was found that}
\begin{equation}
  \Delta \gamma_e = a \left(\frac{n_u}{10^{20}}\right)^{b} (T_{e,u})^{c} (T_{e,t})^{d} (q_{in})^{e},
  \label{eq:gamma_e_fit2}
\end{equation}
{with}
\begin{equation*}
  a = 3.768,\quad b = -1.123,\quad c = -1.603 ,\quad d = -0.886,\quad e = 1.556,
\end{equation*}
{is a good fit to the data, with an error of 0.10. However, it was found that this predicts very large values of $\Delta \gamma_e$ when upstream collisionality is small but temperature gradients are large. As such, it is recommended to use (\mbox{\ref{eq:gamma_e_fit}}) in regimes not studied here, despite slightly worse agreement with the SOL-KiT data. It is also worth noting that no relationship of this kind was found to perform better for $f_{\kappa_e}$ than (\mbox{\ref{eq:f_kappa_fit}}).}

The scalings presented here are straightforward to implement in a fluid model. Because of the spatial variation in heat flux suppression observed in these simulations (see Figure \ref{fig:flux_suppressionc}), implementing the line-averaged quantity $f_{\kappa_e}$ as a prefactor to the Spitzer-H\"{a}rm conductivity may not yield accurate temperature profiles, but should be adequate for predicting the overall power balance {and target temperatures}. These scalings may also be used in simple analytical SOL models such as the modified two-point model \cite{Stangeby2018} or the Lengyel model for predicting detachment onset with radiating impurities \cite{Lengyel1981}.

\begin{figure*}[h]
  \begin{subfigure}[h]{.45\textwidth}
      \centering
      \includegraphics[width=\linewidth]{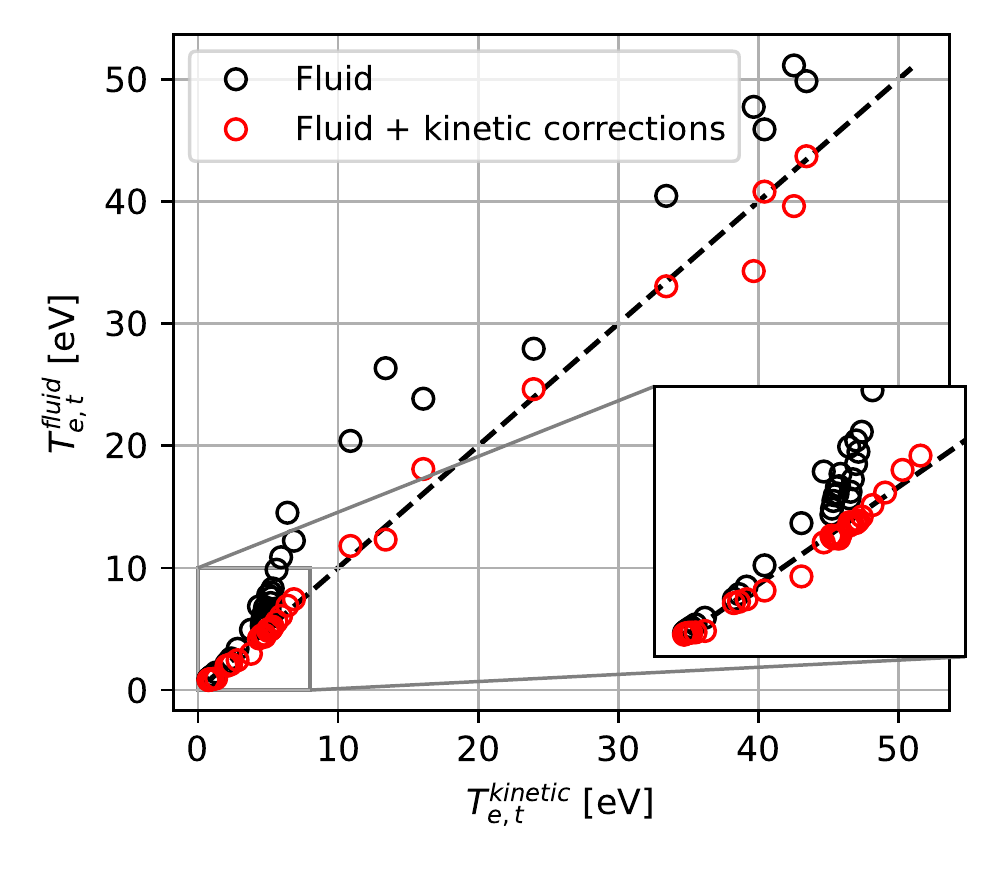}
      \caption{Target electron temperatures in kinetic simulations (x-axis) compared with fluid / fluid with kinetic corrections.}
  \label{fig:T_t_comparison}
  \end{subfigure}
  \hfill
  \begin{subfigure}[h]{.45\textwidth}
      \centering
      \includegraphics[width=\linewidth]{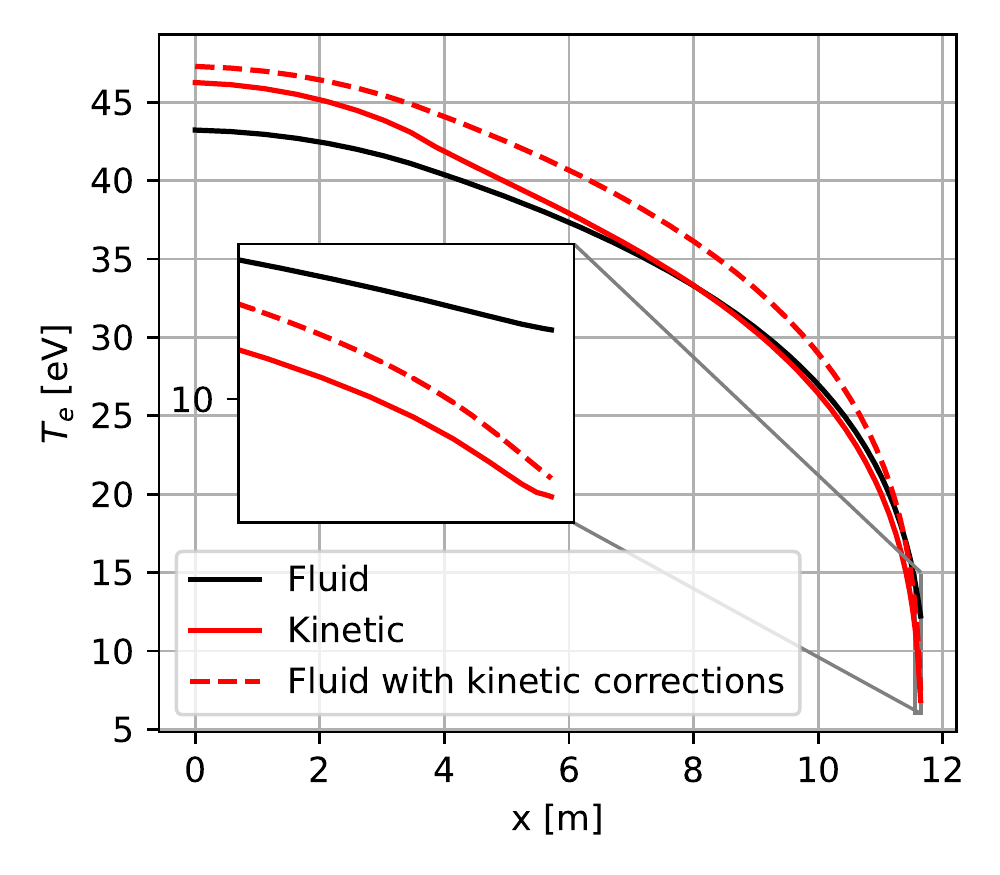}
      \caption{Fluid, kinetic and fluid with kinetic corrections temperature profiles for {a simulation with $q_{in}=64$MWm$^{-2}$, $Ln_u=0.3\times10^{21}$m$^{-2}$.}
      }
  \label{fig:kin_corr_temperature_profiles}
  \end{subfigure}
  \caption{Adding kinetic corrections to SOL-KiT fluid mode using equations (\ref{eq:gamma_e_fit}) and (\ref{eq:f_kappa_fit}).}
  \label{fig:kin_corr}
\end{figure*}

To test these relationships, we have implemented (\ref{eq:f_kappa_fit}) and (\ref{eq:gamma_e_fit}) in the fluid version of SOL-KiT, using self-consistent values of $\nu_{e,u}^*$, to calculate the modifications to $\gamma_e=\gamma_e^{fluid} + \Delta \gamma_e$ and $q_{\parallel,e} = f_{\kappa_e}q_{\parallel,e}^{SH}$. In Figure \ref{fig:T_t_comparison}, we compare the target temperatures in kinetic simulations with those in fluid simulations, with and without the kinetic corrections. Temperature profiles for a particular simulation are shown in Figure \ref{fig:kin_corr_temperature_profiles}. Agreement with $T_{e,t}$ is good, with the RMS error reduced from {43.3\%} to {8.2\%} with the addition of the kinetic corrections. Temperature profiles show that agreement is improved at the upstream and target locations, but differences exist in the rest of the domain {and the $T_e$ uptick close to the wall is not captured}. This is expected due to the line-averaging of $f_{\kappa_e}$.



\begin{figure*}[h]
  \begin{subfigure}[h]{.47\textwidth}
  \centering
      \includegraphics[width=\linewidth]{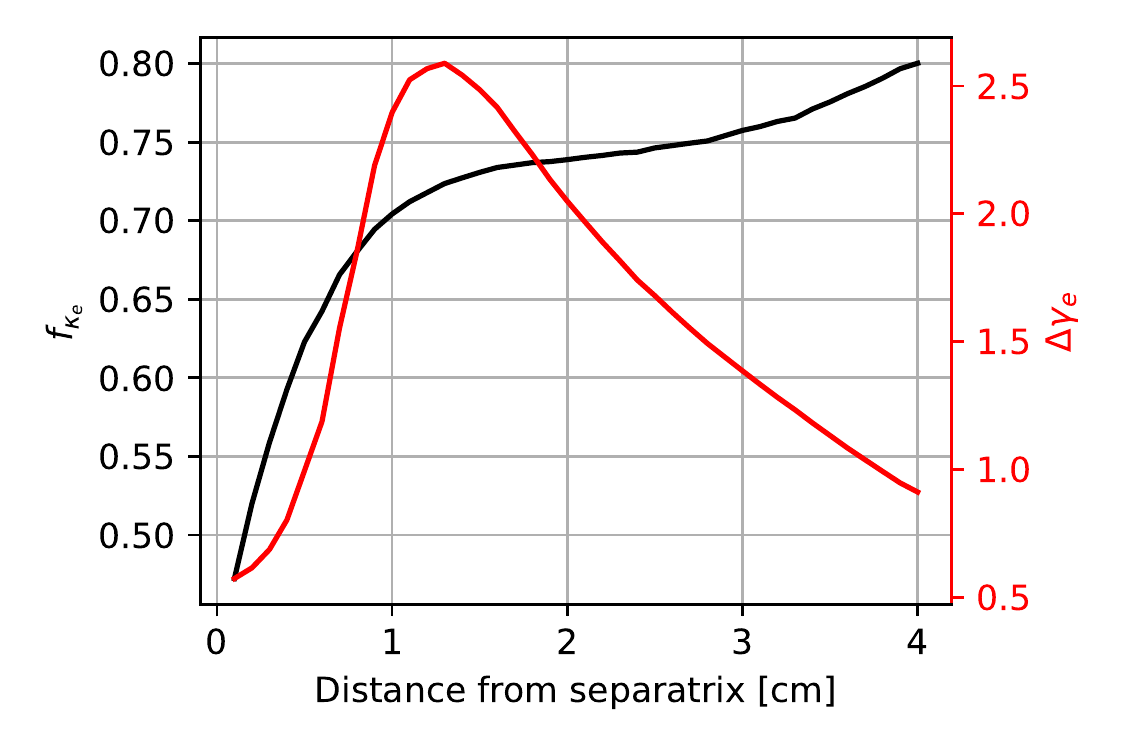}
      \caption{}
      \label{fig:ITER_kinfacs}
  \end{subfigure}
  \hfill
  \begin{subfigure}[h]{.47\textwidth}
      \centering
      \includegraphics[width=\linewidth]{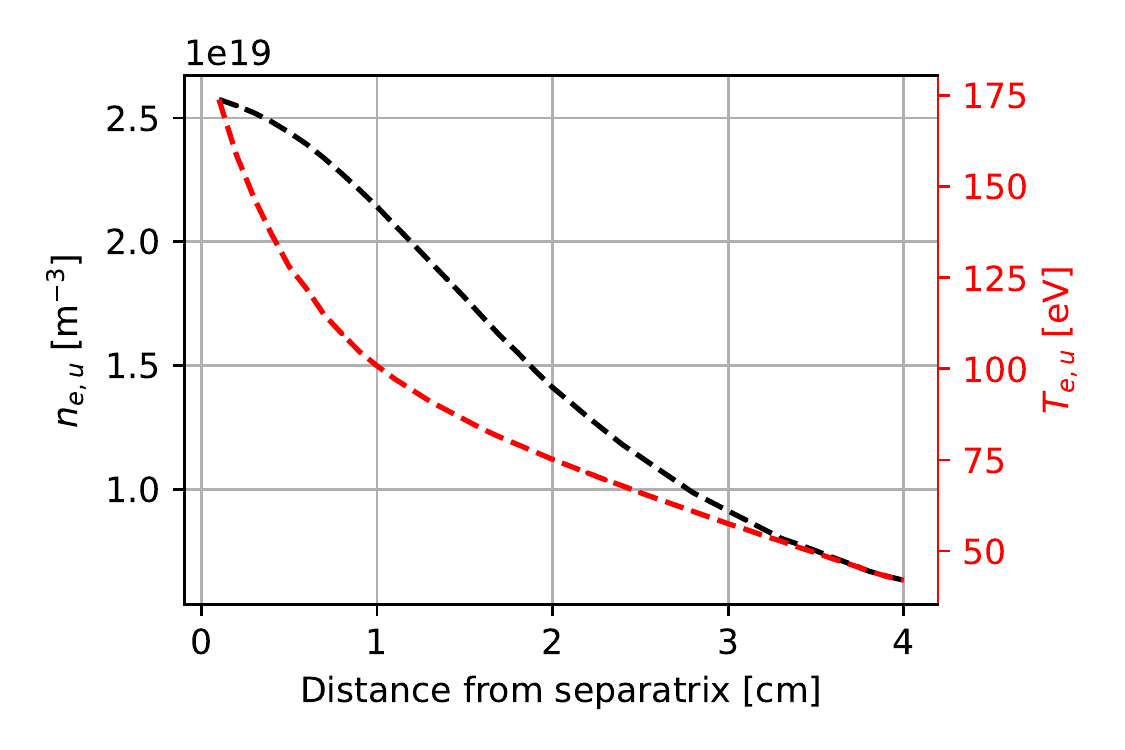}
      \caption{}
      \label{fig:ITER_radprofiles}
  \end{subfigure}
  \caption{{Radial profiles at the outer midplane of (a) kinetic factors and (b) plasma profiles for the ITER scenario modelled in \mbox{\cite{Veselova2021}}.}
    }
  \label{fig:ITER_test}
\end{figure*}

If we assume these scalings can be applied for larger values of {$T_{e,u}$ and $Ln_u$, then for the plasma profiles obtained with the ITER scenario modelled in \mbox{\cite{Veselova2021}} (`standard transport' case; profiles shown in Figure \mbox{\ref{fig:ITER_radprofiles}}), Figure \mbox{\ref{fig:ITER_kinfacs}} shows the expected values of kinetic factors as a function of radial distance at the outer midplane. Here, $q_{in}=800$MWm$^{-2}$ at the separatrix has been assumed, with a radial decay width of $\lambda_q=3$mm. The collisionality ranges from $
\nu_{e,u}^*\sim6-25$. We see values of $\Delta \gamma_e$ up to 2.59 at the peak, representing an enhancement of around 50\% over the classical value, and over 50\% heat flux suppression just beyond the separatrix.  It would be straightforward to implement (\mbox{\ref{eq:f_kappa_fit}}) and (\mbox{\ref{eq:gamma_e_fit}}) in a code such as SOLPS-ITER to explore the significance of such effects. 
}

As an additional test of how these scaling relationships behave in reactor relevant conditions, the scaling equations were implemented into a two-point model (2PM) \cite{Stangeby2001} numerical solver, with $f_{\kappa_e}$ and $\Delta \gamma_e$ scalings incorporated as modifications to the $\kappa_e$ and $\gamma$ terms. This modified 2PM is solved iteratively to find converged self-consistent solutions with the kinetic factors, and applied to SPARC-like SOL conditions \cite{Kuang2020} covering a $\nu_{e,u}^*$ range of $\sim$0.2-100. The low collisionality limiting behaviour for the scalings described above was observed with this tool. Significantly steeper temperature gradients, with higher $T_{e,u}$ and lower $T_{e,t}$, were observed in the 2PM solutions at intermediate and low collisionality, as a result of the added $f_{\kappa_e}$ and $\Delta \gamma_e$ factors. Full results of this 2PM study will be published elsewhere.


A caveat to {the fits presented} is that the plasma model in SOL-KiT does not currently include flux tube expansion (or other SOL geometry effects) or contributions from molecules or impurity species. The former will 
alter 
{plasma behaviour}, {in particular with respect to detachment, and there may also be a kinetic effect due to the mirror force on the electrons.} The latter will represent additional particle, momentum and energy sources/sinks. An ongoing project to redevelop SOL-KiT with a more flexible physics model, as well as improved computational efficiency and parallelisation, should make it possible to study kinetic effects in the presence of such additional physics.

\section{Conclusion}
\label{sec:conclusion}

We have presented kinetic studies of electron transport in tokamak scrape-off layer plasmas across a range of input powers and densities, under steady-state conditions.

One of the primary aims of this study has been to validate the local approximation in fluid models, which are frequently employed to model SOL plasmas. We see that, for SOL equilibria, a kinetic treatment of the electrons does not change qualitative behaviour in terms of the particle flux to the target with this plasma model, as shown in Figure \ref{fig:rollover_plots}, despite changes to the electron temperature profiles and reductions in the target temperatures (Figure \ref{fig:flux_suppression}). 

Typically, it has been assumed that the classical value of $\gamma_e$ is valid at equilibrium, but here there are differences of up to {98\%}, as shown in Figure \ref{fig:gamma_e_enhancement}. We provide a qualitative understanding of this enhancement in terms of an imprint of the fast electrons from upstream on the electron distribution at the target. The presence of this enhanced tail is predicted to have significant impacts on collision rates for inelastic processes with threshold energies $\gtrsim$ 50eV, for example the ionisation of plasma impurities. This is the subject of an ongoing study.

The enhancement of $\gamma_e$ and reduction in $q_{\parallel,e}$ at equilibrium is shown to follow scalings based on basic SOL parameters, (\ref{eq:f_kappa_fit}) \& (\ref{eq:gamma_e_fit}). The performance of these fits is shown in Figures \ref{fig:f_kappa_fit} and \ref{fig:gamma_e_fit}. To test the ability of these corrections to capture kinetic effects in SOL simulations, we have shown that implementing them in the fluid version of SOL-KiT does improve agreement with the fully kinetic $T_e$ profiles. While there are caveats to the use of these scalings outside of SOL-KiT simulations, particularly in relation to the aspects of SOL physics not included in the model used here, it does suggest it is viable to capture kinetic effects at equilibrium in studies of future devices, either in fluid codes or reduced analytical models. Extrapolating to the ITER tokamak for example does predict significant kinetic effects, suggesting at the very least that further study into non-local parallel transport in reactor-class tokamaks is warranted.

{To the authors' knowledge, this is the first attempt at providing relatively simple scaling laws for kinetic effects in equilibrium SOL plasmas. The principle that kinetic effects are directly related to basic descriptors of SOL conditions is a potentially useful approach to analysing their significance.}

The modifications to $\gamma_e$ and $q_{\parallel,e}$ in conjunction with good agreement in power balance and target particle flux behaviour (discussed in Section \ref{sec:discussion} and shown in Figure \ref{fig:rollover_plots}), suggest that both effects contribute in a way which approximately cancels. As such, attempts to capture kinetic effects in fluid models should treat both phenomena simultaneously.  

In this study, the changes to $\gamma_e$ and $q_{\parallel,e}$ are in contrast to the behaviour at equilibrium found using the PIC code BIT1 in \cite{Tskhakaya2007}. There, $\gamma_e$ is found to be well-approximated by the classical value, and $q_{\parallel,e}$ is a non-monotonic function of collisionality, which is contradicted by (\ref{eq:f_kappa_fit}). There are significant differences in the simulations carried out in \cite{Tskhakaya2007}, in particular that plasma-neutral interactions were neglected and that only attached regimes were studied. 
Furthermore, the differences in $\gamma_e$ seen here are of a similar magnitude to those seen with the KIPP code in \cite{Chankin2018}. 

It should also be noted that this investigation has been done for equilibrium plasma conditions. For the sheath boundary condition in particular, much stronger kinetic effects may be present in transient regimes as shown in \cite{Mijin2020a,Tskhakaya2007}, albeit for short durations relative to inter-ELM equilibria.

\section*{Acknowledgements and supporting data}
This work was part funded by the UK Engineering and Physical Science Research Council (EPSRC) (grant number EP/W006839/1) and the UK Atomic Energy Authority (UKAEA). It has also been improved through informal discussions with colleagues at UKAEA and elsewhere, in particular David Moulton and Mike Kryjak. 

The simulations were carried out using the Imperial College London High Performance Computing resource. The simulation data used in this study, along with the analysis scripts, can be found at \url{https://doi.org/10.14469/hpc/10979}.


\bibliographystyle{unsrt}
\bibliography{library}

\end{document}